\def\draftversion{false}

\RequirePackage{ifthen}
\ifthenelse{\equal{\draftversion}{true}}{
  \documentclass[aps,prl,10pt,galley,amsmath,amssymb]{revtex4}
}{
  \documentclass[aps,prl,10pt,twocolumn,amsmath,amssymb,longbibliography,nofootinbib]{revtex4-1}
}

\usepackage{graphicx}
\usepackage[usenames,dvipsnames]{color} 
\usepackage{bm} 
\usepackage{soul} 
\usepackage{booktabs}
\usepackage{mathrsfs}
\usepackage{float}

\ifthenelse{\equal{\draftversion}{true}}{
  \marginparwidth 2.7in
  \marginparsep 0.5in
  \newcounter{comm} 
  \def\commnext{\stepcounter{comm}}
  \def\commtext{{\bf\color{blue}[\arabic{comm}]}}
  \def\commmar{{\bf\color{blue}[\arabic{comm}]}}
  \def\dvm#1{\commnext\marginpar{\small DV\commmar: #1}\commtext}
  \def\dsm#1{\commnext\marginpar{\small DS\commmar: #1}\commtext}
  \def\mlab#1{\marginpar{\small\bf #1}}
  
}{
  \def\dvm#1{}
  \def\dsm#1{}
  \def\mlab#1{}
  
}

\newcommand{\beq}{\begin{equation}}
\newcommand{\eeq}{\end{equation}}
\newcommand{\bea}{\begin{eqnarray}}
\newcommand{\eea}{\end{eqnarray}}

\newcommand{\eq}[1]{Eq.~(\ref{eq:#1})}

\newcommand{\eqs}[2]{Eqs.~(\ref{eq:#1}) and (\ref{eq:#2})}

\newcommand{\eqo}[2]{Eq.~(\ref{eq:#1}) or (\ref{eq:#2})}
\newcommand{\eqr}[2]{Eqs.~(\ref{eq:#1}-\ref{eq:#2})}
\newcommand{\me}[3]{\langle#1\vert#2\vert#3\rangle}
\newcommand{\ev}[1]{\langle#1\rangle}
\def\z2{$\mathbb{Z}_2$}

\def\rr{{\bf r}}

\renewcommand{\say}[1]{``#1''}

\DeclareMathAlphabet\mathbfcal{OMS}{cmsy}{b}{n}   

\def\mq{Q}              
\def\mqb{\mathbf{Q}}    
\def\mqd{\mathcal{Q}}  
\def\mqdb{\bm{\mqd}}    

\begin{document}


\title{Towards a theory of surface orbital magnetization}

\author{Daniel Seleznev}
\affiliation{
Department of Physics and Astronomy, Center for Materials Theory, Rutgers University,
Piscataway, New Jersey 08854, USA}

\author{David Vanderbilt}
\affiliation{
Department of Physics and Astronomy, Center for Materials Theory, Rutgers University,
Piscataway, New Jersey 08854, USA}

\date{\today}
\begin{abstract}

The theory of bulk orbital magnetization has been formulated both in reciprocal space based on Berry curvature and related quantities, and in real space in terms of the spatial average of a quantum mechanical local marker. Here we consider a three-dimensional antiferromagnetic material having a vanishing bulk but a nonzero surface orbital magnetization.  We ask whether the surface-normal component of the surface magnetization is well defined, and if so, how to compute it.  As the physical observable corresponding to this quantity, we identify the macroscopic current running along a hinge shared by two facets. However, the hinge current only constrains the difference of the surface magnetizations on the adjoined facets, leaving a potential ambiguity.  By performing a symmetry analysis, we find that only crystals exhibiting a pseudoscalar symmetry admit well-defined magnetizations at their surfaces at the classical level.  We then explore the possibility of computing surface magnetization via a coarse-graining procedure applied to a quantum local marker.  We show that multiple expressions for the local marker exist, and apply constraints to filter out potentially meaningful candidates.  Using several tight-binding models as our theoretical test bed and several potential markers, we compute surface magnetizations for slab geometries and compare their predictions with explicit calculations of the macroscopic hinge currents of rod geometries. We find that only a particular form of the marker consistently predicts the correct hinge currents.

\end{abstract}

\maketitle


\section{Introduction}
\label{Intro}

The modern theories of  bulk electric polarization $\textbf{P}$ and bulk
orbital magnetization $\textbf{M}$ \cite{king-smith-prb1993,vanderbilt-prb1993,xiao-prl2005,thonhauser-prl2005, ceresoli-prb2006,shi-prl2007,souza-prb2008}
express these quantities as Brillouin zone integrals of quantities
involving Berry connections and curvatures of ground state Bloch
functions. However, the ground-state of systems of independent electrons
may also be uniquely described by the single-particle density matrix, also
known as the ground state projector $P(\textbf{r},\textbf{r}')$. This is a
quantity that for insulating states of matter decays exponentially with
$|\textbf{r}-\textbf{r}'|$, even for Chern insulators
\cite{thonhauser-prb2006}.

A natural question to ask is whether it is possible to express
$\textbf{P}$ and $\textbf{M}$ in terms of $P(\textbf{r},\textbf{r}')$.
In the case of the polarization, while the change
$\Delta\textbf{P}$ can be determined by following the change of
$P(\textbf{r},\textbf{r}')$ between two states connected by an
adiabatic switching process, there is no corresponding expression
for $\textbf{P}$ itself, which is
determined only modulo a quantum \cite{king-smith-prb1993,vanderbilt-prb1993}.

However, $\textbf{M}$ does not suffer from any quantum of indeterminacy,
and should be expressible via the ground state projector. Bianco and
Resta \cite{bianco-prl2013} demonstrated that this is, in fact, correct.
Starting from an expression for a simple 2D crystallite of finite area,
the authors demonstrated that for any insulator, even a Chern insulator,
the orbital magnetization $M$ is given in the thermodynamic limit by
\begin{equation}
M=\frac{1}{A}\int_A\mathcal{M}(\textbf{r})d\textbf{r}.\label{eq:MarkerEq}
\end{equation}
Here $A$ is the bulk unit cell area and $\mathcal{M}(\textbf{r})$ is a local function, hereafter referred to as the local marker, which is expressed in terms of $P(\textbf{r},\textbf{r}')$. For large finite samples in the thermodynamic limit, the averaging in Eq.~(\ref{eq:MarkerEq}) may equivalently be performed over the entire crystallite. Although $\mathcal{M}(\textbf{r})$ appears to play the role of magnetic dipole density, \textit{only it's macroscopic average bears any physical meaning}. This is in keeping with the fact that while microscopic charge and current densities are well defined, microscopic dipole densities (be it charge or orbital magnetic) are not \cite{hirst-rmp1997}.

Recent years have witnessed numerous efforts aimed at
defining and providing explicit expressions for surface-specific
analogues of bulk quantities. Examples include the
surface anomalous Hall conductivity (AHC)
\cite{rauch-prb2018,varnava-prb2020}
and the surface-parallel boundary electric polarization \cite{zhou-prb15,benalcazar-sa17,benalcazar-prb2017,ren-prb2021,trifunovic-prresearch2020} on the surface of a bulk insulator.
To date, however, there has been very little discussion of surface
orbital magnetization in the literature \cite{zhu-prb2021,bianco-prb2016}, and a complete theory of orbital magnetization at the surface of a bulk material has yet to be developed.

In the present work, we explore the
possibility of defining surface orbital magnetization.
By this we mean the excess surface-normal macroscopic magnetization (magnetic moment
per unit area) at the surface of a bulk material with broken
time-reversal (TR) symmetry.
Our investigation is restricted to antiferromagnetic systems
featuring a vanishing bulk
magnetization, as this allows us to readily disentangle surface
contributions to the orbital magnetization from those stemming from
the bulk.  We shall also restrict ourselves throughout to the case
of insulating surfaces of insulating crystals, leaving aside for now
any special considerations that might arise for metallic surfaces.

The case of surface spin magnetization is much more straightforward,
as the latter can be obtained simply by integrating the net spin density
$\Delta n=n_\uparrow-n_\downarrow$ over the surface region after an
appropriate coarse-graining (e.g., using window-averaging methods).
In the remainder of this work, therefore, we focus solely on the orbital
component of the surface magnetization.  Here it is natural to
expect more difficulties, since even the theory of bulk orbital
magnetization has been put on a solid footing only in relatively recent
times \cite{xiao-prl2005,thonhauser-prl2005,ceresoli-prb2006,shi-prl2007,souza-prb2008}.
The essential problem, as in the theory of electric polarization,
is that the position operator is ill defined in the Bloch representation,
so that methods based on Berry connections and curvatures are
required instead.  Another way to frame the problem is to note that
quantum-mechanical
expressions are available for neither
the local polarization density $\mathbf{P}(\mathbf{r})$
nor the local orbital magnetization density $\mathbf{M}(\mathbf{r})$.

Even at a classical level,
another difficulty arises. Recall that, while the bulk
orbital magnetization cannot be inferred from the local current
distribution deep in the bulk, it can be deduced with the added
knowledge of the currents on its surface facets.
By analogy, one might expect
that the surface magnetization on a given facet can be
inferred in a similar way from the added knowledge of the currents
flowing at the edges of the facet. However, such a facet boundary is
always a hinge where two facets meet, with the hinge
current given by the \textit{difference} between the  surface magnetizations
on the two adjoining facets.  Since the hinge
currents only determine surface magnetization differences, this
raises the question whether surface orbital magnetization can be
uniquely determined, or only determined up to a constant shift in the
values predicted for all facets, even when given a perfect knowledge
of all local currents.  Such a ``shift freedom'' is an unavoidable
feature of a classical description starting from current densities,
and is related to the fact that while the curl of $\mathbf{M}(\mathbf{r})$ is
constrained by the current distribution, its divergence is not.

With the introduction of a quantum description, the bulk
electric polarization and orbital magnetization, which were
ill determined from a classical knowledge of bulk charge and current
densities, become well defined based on a knowledge of the bulk
Bloch eigenstates (up to a quantum in
the case of the polarization).  In a similar way, one might hope that
an appropriate quantum description of the surface problem would
allow for a robust prescription for computing surface orbital
magnetization from a knowledge of surface, as well as bulk,
ground state wave functions. This is the goal of the present work.

In this paper, we first investigate the role of symmetry and
show that certain classes of symmetries do allow for the surface
magnetization to be uniquely extracted from a
knowledge of hinge currents, even at the classical level.
We then turn to the quantum problem and explore
whether the use of a local marker $\mathcal{M}(\textbf{r})$, such
as the one proposed by Bianco and Resta, can be adopted for the
purpose of defining a surface orbital magnetization.  That is,
even if the value of $\mathcal{M}(\textbf{r})$ at a single point
has no obvious physical meaning, we ask whether it is possible to
coarse-grain and integrate such a marker over the surface region
to obtain a valid expression for the surface orbital magnetization.

This paper is organized as follows. In Sec.~\ref{Qualitative}, we
identify the physical observable corresponding to the presence of a
surface magnetization, namely the macroscopic current running along a
hinge shared by adjacent surface facets of a bulk crystal,
given by the difference between the
magnetizations of the two surface facets forming the hinge.
We show that within the framework of a classical theory,
this single relation is insufficient to ascribe a uniquely defined 
value of magnetization to any of the facets; constraints in the form of
crystalline symmetries are required to accomplish this. In
Sec.~\ref{Symms}, we identify all the possible symmetries that lead to
unambiguous surface magnetizations. We find that this set of symmetries is identical to the set of symmetries that quantizes the Chern-Simons axion coupling.

In Sec.~\ref{Methods} we introduce in
greater detail the local marker formulation of orbital magnetism, and
introduce our formalism for the calculation of surface magnetization and
hinge currents. The results of Sec.~\ref{Symms} and Sec.~\ref{Methods}
are later tested in Sec.~\ref{Results} by calculations performed on
tight-binding models. We end our paper with a discussion of our
results and some speculations on their connection to the theory of orbital
magnetic quadrupole moments in Sec.~\ref{Discussion}, and summarize in
Sec.~\ref{Summary}.

\section{Surface Magnetizations and Hinge Currents}
\label{Qualitative}

Consider a stand-alone 2D system such as a monolayer or
multilayer with layer-normal magnetization $M_\perp$.
The magnetization manifests itself on any edge of the system as
a macroscopic bound current of magnitude $M_\perp$, with its sign determined from the right-hand rule. At the classical level, $M_\perp$ can be determined from the combined knowledge of the microscopic current density deep in the bulk and at the edge. Without the added knowledge of the current density at the edge, however, $M_\perp$ can be determined only from a knowledge of the quantum-mechanical bulk Bloch eigenstates.

\begin{figure}[t]
\centering \includegraphics[scale=0.45]{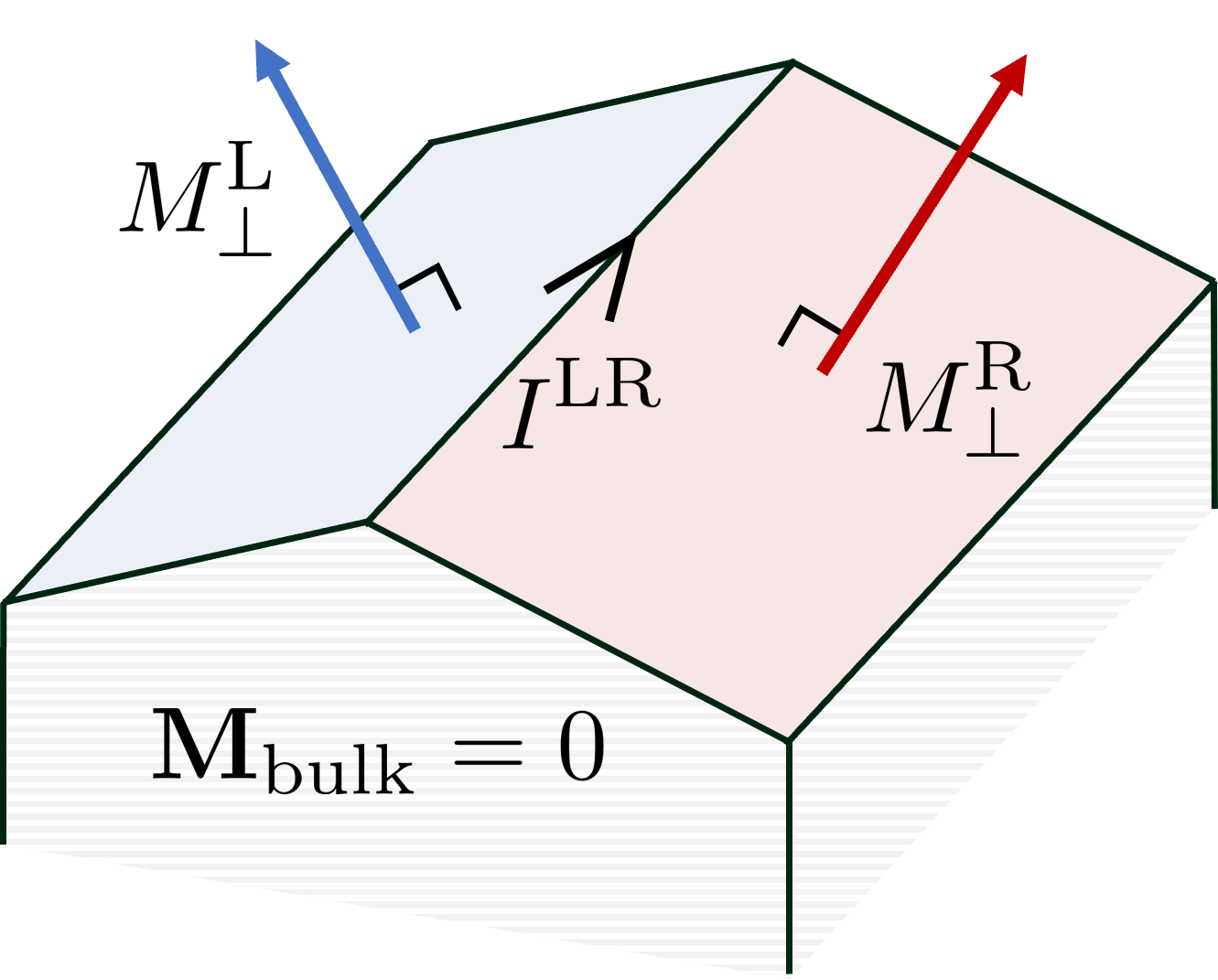}
\caption{Schematic depicting a hinge and the surface facets that compose
it. The black arrowhead on the hinge depicts the direction of flow for the hinge current $I^{\text{LR}}$ to be positive. The maroon and blue arrows denote the surface-normal magnetization vectors $M^{\text{R}}_{\perp}$ and
$M^{\text{L}}_{\perp}$, respectively. The hinge current is $I^{\text{LR}}=M^{\text{L}}_{\perp}-M^{\text{R}}_{\perp}$.
}
\label{fig:facets}
\end{figure}

If instead we consider the surface of a 3D bulk system (with vanishing bulk magnetization), we would also expect any surface-normal magnetization at the surface to manifest itself at an edge in a similar fashion. However, in the presence of the bulk, the objects playing the role of edges are the hinges, i.e., the intersections of neighboring surface facets. So, while in the case of an isolated 2D system the edge current directly determines the magnetization $M_\perp$, any current present on a hinge only determines the \textit{difference} of $M_\perp$ values
on the two surface facets meeting at the hinge. Hence a classical knowledge of the microscopic current distribution in the bulk, at the surfaces, and at the hinge is sufficient to uniquely determine the difference of $M_\perp$ values across facets sharing a hinge, but not the $M_\perp$ values individually. We depict this
situation schematically in Fig.~\ref{fig:facets}, where it
is evident that

the hinge current $I^{\text{LR}}$ is given by
\begin{equation}
I^{\text{LR}}=M^{\text{L}}_{\perp}-M^{\text{R}}_{\perp}.
\label{eq:MainHingeEq}
\end{equation}

Note that a line current may also be present at a line
defect, such as a step or domain wall \cite{varnava-natcom2021},
that separates surface patches having
the same orientation but different $M_\perp$ values.
We can regard these as 180$^\circ$ hinges,
and the treatment of such cases introduces no new difficulties.
In the remainder of this work, we restrict the discussion to true hinges
connecting facets of different orientation.
In any case, we emphasize
that the line currents in question are bound currents localized at
insulating hinges, not the chiral conductance channels that may
occur in topological systems, e.g., at the boundary of a 2D quantum
Hall state or on the hinges of an axion insulator~\cite{khalaf-prb2018,varnava-prb2018}.

For the entirety of this paper, we take the
hinge current to be a physical observable, being detectable in
principle by the circulating magnetic field it creates.
Combined with Eq.~(\ref{eq:MainHingeEq}), this
assumption indicates that \textit{differences} between magnetizations on
neighboring surface facets are also observables, in addition to being classically well defined in the sense described earlier. However, there is no \textit{a priori} reason to believe that the individual values of facet
magnetizations are either observables or are uniquely defined in the context of classical theory.

As mentioned in the Introduction and explicitly demonstrated by Eq.~(\ref{eq:MainHingeEq}), the individual values of the facet magnetizations are defined classically only up to a common constant shift, resulting in a shift freedom. To understand how the shift freedom arises, consider a finite crystallite embedded in vacuum. The crystallite's steady state microscopic current density $\textbf{j}(\textbf{r})$ is divergenceless, and thus may be expressed as the curl of a vector field $\textbf{M}(\textbf{r})$ with the interpretation of a local magnetization density.
However, $\textbf{M}(\textbf{r})$ has a ``gauge freedom'' in that augmenting
$\textbf{M}(\textbf{r})$ by
the gradient of an arbitrary scalar field $f(\textbf{r})$ leaves $\textbf{j}(\textbf{r})$ unchanged.
In particular, let $f(\textbf{r})$ be some function that is periodic with
average value $\Delta$ in the interior, vanishes in the vacuum region
outside, and transitions from these behaviors over a
few lattice constants at the surface. Then at the coarse-grained
level, the new $\textbf{M}(\textbf{r})$ differs from the old one by what appears
to be a delta-function concentration of $M_\perp$ of magnitude $-\Delta$ on all
surface facets.  This is precisely the shift freedom.

We must therefore understand whether,
or under what circumstances, we can resolve this ambiguity and assign a
unique $M_{\perp}$ to a given facet of the bulk. Once an unambiguous
value of magnetization is assigned to even a single facet, we may then
repeatedly employ Eq.~(\ref{eq:MainHingeEq}) to find the magnetization
at any other facet of the 3D crystallite.

\begin{figure}[t]
\centering
\includegraphics[width=0.8\columnwidth]{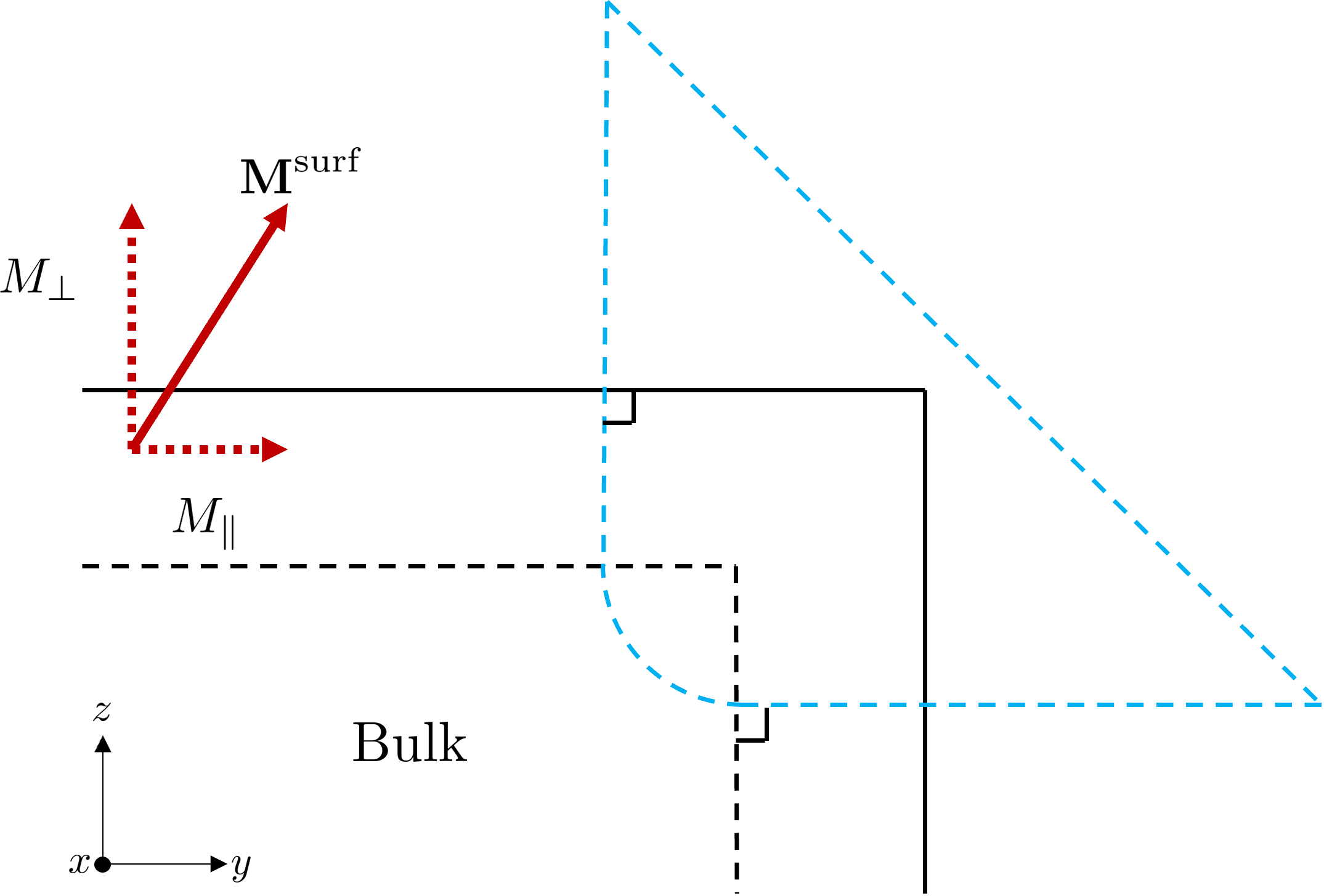}
\caption{Schematic depiction of a hinge formed by two surface
facets, one of which has a magnetization $\textbf{M}^{\text{surf}}$
that is not normal to the surface, having components
$M_{\perp}$ along $\hat{\textbf{z}}$ and $M_{\parallel}$ along $\hat{\textbf{y}}$.
$M_{\parallel}$ is observable in principle by the presence of a
concentration of  magnetic field $B_y$ in the surface region. The
hinge current is observable via the integral $\oint\textbf{M}\cdot
d\textbf{l}$ around the dashed blue Amperian loop, which remains
normal to the surface while passing through the surface region
delineated by the dashed black line.}
\label{fig:loops}
\end{figure}

While we have focused above on the surface-normal component
of the surface orbital magnetization, an in-plane component
$M_\parallel$ could also be present.  If so, its strength
is easily computed from a detailed knowledge of the microscopic
current distribution in the surface region, unlike the more
problematic $M_\perp$.
In principle, $M_\parallel$ should be observable by
the presence of a corresponding concentration of $B_\parallel$
in the surface region (since $\textbf{H}$ vanishes in this geometry in the
absence of free current).
When $M_\parallel$ is present,
we can filter out its contribution to the hinge current by constructing
an appropriate Amperian loop that is
normal to both surface facets, as depicted in Fig.~\ref{fig:loops}.
Integrating the magnetization around this loop yields the bound current
passing through the loop. Since we assume the absence of free current,
and since the vacuum and bulk are free of magnetization,
only the top and side $M_{\perp}$ values contribute to the hinge current.
We adopt this as the definition of the hinge current in
such situations.

\section{Symmetries and Surface Magnetization}
\label{Symms}

The goal of this section is to establish under what circumstances we can specify
a uniquely defined magnetization on a given surface facet of a 3D crystallite in the context of classical theory. By this, we mean whether a classical knowledge of the microscopic current distribution in the bulk, at the surfaces, and at the hinges is sufficient to assign a unique value of $M_{\perp}$ to a given surface facet. As noted in the previous section, Eq.~(\ref{eq:MainHingeEq}) effectively results in one equation in two unknowns, which prevents an assignment of definite values to $M^{\text{L}}_{\perp}$ and $M^{\text{R}}_{\perp}$. We therefore need additional restrictions on the possible values $M^{\text{L}}_{\perp}$ and $M^{\text{R}}_{\perp}$ may take. In this Section, these restrictions will appear in the form of symmetry-induced constraints. We first introduce the notation we will be using to describe the actions of the symmetry operations, and subsequently delve into the symmetry analysis of surface magnetization.

\begin{table}[b]
\begin{ruledtabular}
\begin{tabular}{lcc}
  \text{ }& $M_{\perp}$-preserving & $M_{\perp}$-reversing\\ [0.5ex]           
\colrule
$\hat{\textbf{n}}$-reversing &$R^+_{\hat{\textbf{n}}}$ &$R^-_{\hat{\textbf{n}}}$  \\ [0.5ex]
$\hat{\textbf{n}}$-symmorphic & $S^+_{\hat{\textbf{n}}}$ & $S^-_{\hat{\textbf{n}}}$ \\ [0.5ex]
$\hat{\textbf{n}}$-nonsymmorphic & $N^+_{\hat{\textbf{n}}}$ & $N^-_{\hat{\textbf{n}}}$ \\ [0.5ex]
\end{tabular}
\end{ruledtabular}
\caption{\label{tab:SymmLabels}
Summary of labels attached to a symmetry operation depending on whether it reverses $\hat{\textbf{n}}$ ($R$) or not ($S,N$), where $S$ and $N$ refer to $\hat{\textbf{n}}$-symmorphic and $\hat{\textbf{n}}$-nonsymmorphic operations respectively. Superscripts `$-$' and `$+$' indicate whether or not the operation reverses the sign of the outward-directed $M_\perp$ on a surface with normal $\hat{\textbf{n}}$, which also corresponds to whether the symmetry operation is axion-odd or axion-even.}
\end{table}

\subsection{Notation} 

Consider a surface with unit normal $\hat{\textbf{n}}$. Let the outward surface-normal component of the surface magnetization be $M_\perp=\textbf{M}\cdot\hat{\textbf{n}}$,
with units of magnetic moment per unit area. With respect to this choice of $\hat{\textbf{n}}$,
we will focus on operations that either preserve or reverse the direction $\hat{\textbf{n}}$, as
well as those that preserve or flip the sign of $M_{\perp}$. Operations
that flip $\hat{\textbf{n}}$ will be denoted by
$R^{\pm}_{\hat{\textbf{n}}}$; the superscript `$+$' or `$-$' indicates
that $M_{\perp}$ is preserved or flipped, respectively.
We further classify operations that preserve $\hat{\textbf{n}}$ as
either symmorphic or nonsymmorphic along that direction, where by the latter term we mean operations that involve a fractional translation along $\hat{\textbf{n}}$, such as a screw or glide mirror operation.
Thus, $\hat{\textbf{n}}$-symmorphic operations will be denoted
$S^{\pm}_{\hat{\textbf{n}}}$
(even if they involve fractional translations in the plane
normal to $\hat{\textbf{n}}$),
while $\hat{\textbf{n}}$-nonsymmorphic
operations will be denoted $N^{\pm}_{\hat{\textbf{n}}}$.

We summarize the classification of symmetries in
Table \ref{tab:SymmLabels}. We note that operations labeled with a superscript `$-$' are axion-odd while those labeled with a `$+$' superscript are axion-even.
By definition, an axion-odd symmetry is either a
proper spatial rotation with time reversal or
an improper rotation without time reversal. The significance of this distinction will become clear in Sec.~\ref{sec:discussion}.

For concreteness, let $\hat{\textbf{n}}=\hat{\textbf{z}}$. We list all
the symmetry operations that fall into the categories listed above in
Table~\ref{tab:Symms}.
In our notation,
$E$ denotes the identity operation and $I$ denotes inversion.
$C_n$ denotes a rotation by $2\pi/n$ about the $\hat{\textbf{z}}$ axis
($n=2,3,4,6$) while $\overline{C}_2$ is a 2-fold rotation about
an axis in the $x$-$y$ plane. 
$m_z$ denotes a mirror reflection across the $x$-$y$ plane, and
$m_d$ is a mirror reflection across a plane containing the
$\hat{\textbf{z}}$ axis. 
$S_n=m_zC_n$ denotes an improper rotation (rotoinversion) about the $\hat{\textbf{z}}$ axis. 
A prime indicates composition with time reversal.
Nonsymmorphic fractional translations along $\hat{\textbf{z}}$ are
indicated as $c/2$ or $pc/n$,
where $c$ is the vertical lattice constant and $p<n$ is an integer
(or half-integer in the case of $C'_3$).
Half lattice translations in the $x$-$y$ plane are indicated by
$\boldsymbol{\tau}_{d/2}$. In what follows, we shall often denote
the screw operation $\{C_n\vert pc/n\}$ as just $n_p$ for conciseness.

It is important to note that some of the above
operations may belong to $R$, $S$, and $N$ operations with respect to
other directions.  For example, the glide mirror $\{m_y|c/2\}$ is both
an $N^-_{\hat{\textbf{z}}}$ and an $R^-_{\hat{\textbf{y}}}$ operation.

\begin{table}[b]
\begin{ruledtabular}
\begin{tabular}{lc}
  Type&Symmetry Operations\\ [0.5ex]           
\colrule
$R^-_{\hat{\textbf{z}}}$ & $m_z$, $I$, $S_{3,4,6}$,
$\overline{C}^{\prime}_2$, $\{m_z|\boldsymbol{\tau}_{d/2}\}$,
$\{\overline{C}^{\prime}_2|\boldsymbol{\tau}_{d/2}\}$  \\ [0.5ex]

$S^-_{\hat{\textbf{z}}}$ & $E^{\prime}$, $\{E'|\boldsymbol{\tau}_{d/2}\}$,  $C^{\prime}_{n}$, $m_d$ \\ [0.5ex]

$N^-_{\hat{\textbf{z}}}$ & $\{E^{\prime}|c/2\}$, $\{C^{\prime}_{n}|pc/n\}$, $\{m_d|c/2\}$
 \\ [0.5ex]

$R^+_{\hat{\textbf{z}}}$ & $m^{\prime}_z$, $I^{\prime}$, $S^{\prime}_{3,4,6}$, $\overline{C}_2$, $\{m^{\prime}_z|\boldsymbol{\tau}_{d/2}\}$,
$\{\overline{C}_2|\boldsymbol{\tau}_{d/2}\}$  \\ [0.5ex]

$S^+_{\hat{\textbf{z}}}$ & $C_{n}$, $m^{\prime}_d$ \\ [0.5ex]

$N^+_{\hat{\textbf{z}}}$ & $\{C_{n}|pc/n\}$, $\{m^{\prime}_d|c/2\}$ \\
[0.5ex]
\end{tabular}
\end{ruledtabular}
\caption{\label{tab:Symms}List of all the specific symmetry
operations that belong to each symmetry operation type.}
\end{table}

\subsection{Symmetry Analysis for Surface Magnetization}
\label{sec:sym-analysis}

In this subsection, we explore how to exploit the symmetry operations in
the $S^{\pm}_{\hat{\textbf{n}}}$, $R^{\pm}_{\hat{\textbf{n}}}$, and
$N^{\pm}_{\hat{\textbf{n}}}$ classes to define an unambiguous surface
magnetization on at least one surface facet; repeated application
of Eq.~(\ref{eq:MainHingeEq}) will subsequently help define
magnetizations for all other surface facets. The main idea is that a
given symmetry operation will relate specific surface facets to one
another, allowing us to explicitly relate physical quantities defined on
those facets.
We then consider the magnetizations $M_\perp$ at the surfaces,
and check whether this configuration allows for an unambiguous
determination of these $M_\perp$ values. We find that
$S^{-}_{\hat{\textbf{n}}}$, $R^{-}_{\hat{\textbf{n}}}$, and
$N^{-}_{\hat{\textbf{n}}}$ operations allow us to define unambiguous surface
magnetizations, while for $S^{+}_{\hat{\textbf{n}}}$,
$R^{+}_{\hat{\textbf{n}}}$, and $N^{+}_{\hat{\textbf{n}}}$ this is
not possible.

As a reminder, we are focusing on $\hat{\textbf{n}}=\hat{\textbf{z}}$
for concreteness. We first investigate the axion-odd symmetries 
$S^{-}_{\hat{\textbf{z}}}$, $R^{-}_{\hat{\textbf{z}}}$, and $N^{-}_{\hat{\textbf{z}}}$, and subsequently the axion-even symmetries $S^{+}_{\hat{\textbf{z}}}$,
$R^{+}_{\hat{\textbf{z}}}$, and $N^{+}_{\hat{\textbf{z}}}$.

\paragraph{Symmetries of type $S^-_{\hat{\textbf{z}}}$.}

Suppose we find a symmetry of the type $S^-_{\hat{\textbf{z}}}$ in the
bulk symmetry group. We may then construct a surface respecting this
symmetry with unit normal $\hat{\textbf{z}}$. Under an $S^-_{\hat{\textbf{z}}}$
operation, the surface will remain invariant, but the magnetization at the
surface will be flipped. This can only mean that the magnetization at
the surface is zero.

\paragraph{Symmetries of type $R^-_{\hat{\textbf{z}}}$.}

If we find an $R^-_{\hat{\textbf{z}}}$ symmetry in the bulk
symmetry group, we may construct
a finite slab in $z$ consistent with the identified
$R^-_{\hat{\textbf{z}}}$ symmetry applied to the
slab as a whole. This construction will then enforce
identical magnetizations on top and bottom surfaces.

\begin{figure}
\centering
\includegraphics[width=2.5in]{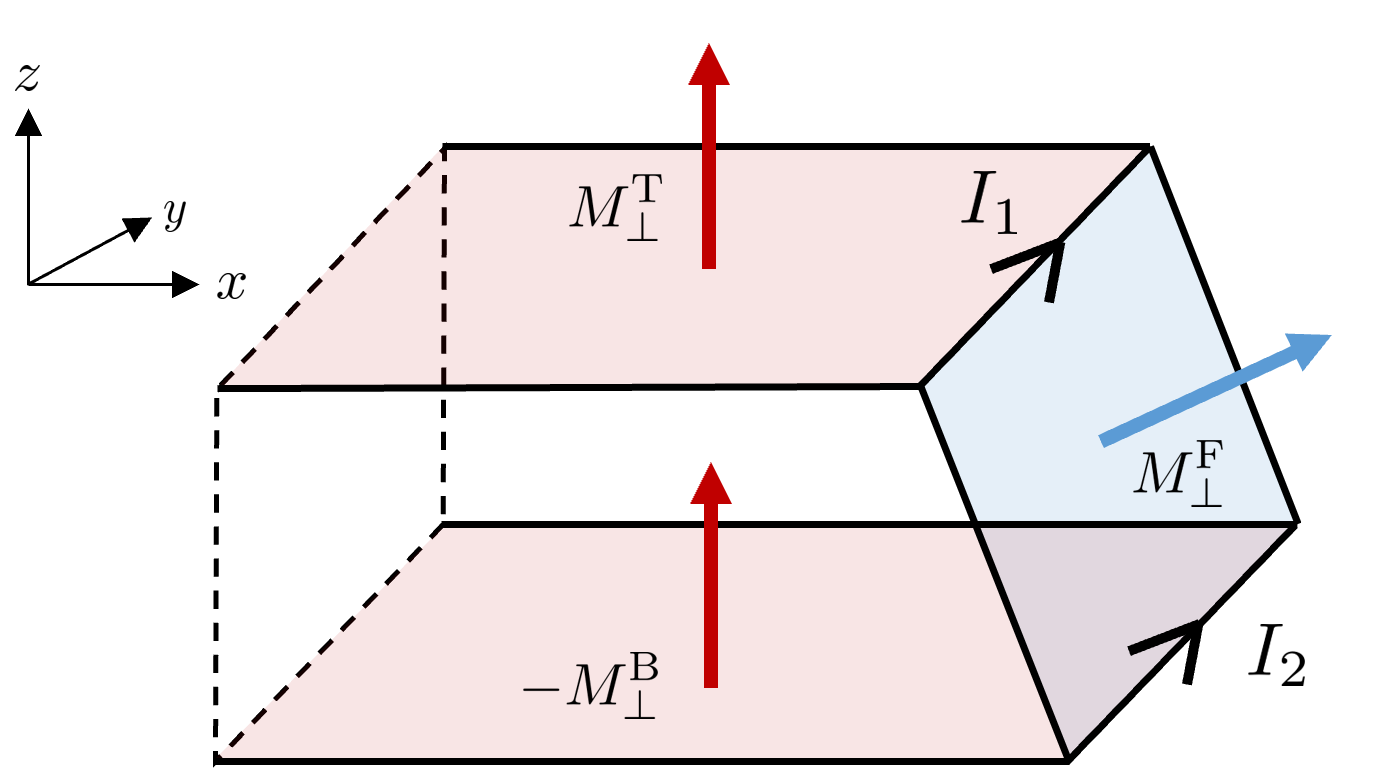}
\caption{Schematic of a slab geometry
for a bulk material with an $R^-_{\hat{\textbf{z}}}$ symmetry.
An additional facet of arbitrary orientation appears at right.
The $R^-_{\hat{\textbf{z}}}$ symmetry ensures that
the magnetizations on the top and bottom facets are equal and parallel, as indicated by the maroon arrows.
All three facet magnetizations may be found using the known hinge currents $I_1$ and $I_2$, as discussed in the text.}
\label{fig:SlabFacets}
\end{figure}

Let us introduce an arbitrary third surface facet to the slab, as shown
in Fig.~\ref{fig:SlabFacets}. In the figure, the arrows normal to the surfaces indicate the direction of the surface-normal magnetization in the given global Cartesian frame. The vectors are labeled by their corresponding magnitudes;
since $M^{\text{B}}_{\perp}<0$, the magnitude of the bottom surface magnetization is labeled as $-M^{\text{B}}_{\perp}$. We then also have two hinges with currents
$I_1$ and $I_2$, with the directions of the arrowheads on the hinges indicating the directions of positive current flow.

Using our knowledge of these currents and the symmetry of the bulk slab,
we may then write down a system of equations

\begin{align}
    0&=M^{\text{T}}_{\perp}+M^{\text{B}}_{\perp},\nonumber\\ I_1&=M^{\text{T}}_{\perp}-M^{\text{F}}_{\perp},\\
    I_2&=M^{\text{F}}_{\perp}-M^{\text{B}}_{\perp},\nonumber
\end{align}
that can be solved for the three facet magnetizations.

\paragraph{Symmetries of type $N^-_{\hat{\textbf{z}}}$.}

From Table~\ref{tab:Symms} we see that there are three types of
operations belonging to the $N^-_{\hat{\textbf{z}}}$ class: the time
reversed half-translation $\{E^{\prime}|c/2\}$, glide mirrors
$\{m_d|c/2\}$, and time reversed screw symmetries.
Note that $\{E^{\prime}|c/2\}$ is also a $S^-_{\hat{\textbf{d}}}$ operation,
while $\{m_d|c/2\}$ and $\{C^{\prime}_{2z}|c/2\}$ are also
$R^-_{\hat{\textbf{d}}}$ operations. We therefore already know that we can
define unambiguous surface
magnetizations in those cases. We thus focus on the time-reversed
screws with 3-fold, 4-fold, and 6-fold rotation axes.

\begin{figure}
\centering
\includegraphics[width=\columnwidth]{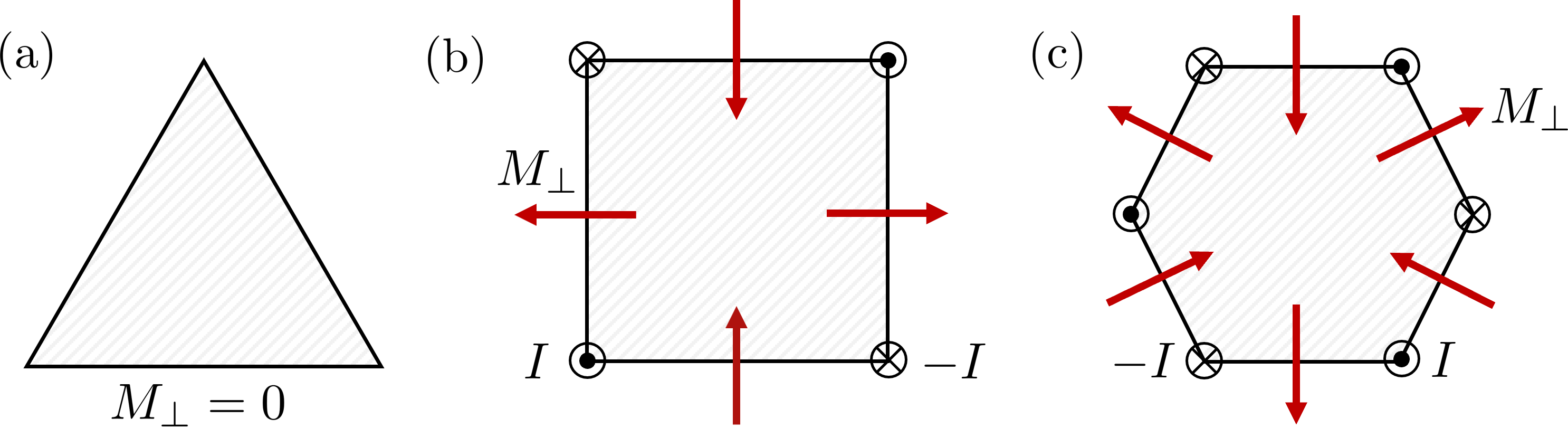}
\caption{
Symmetry constraints arising from $3'_m$, $4'_m$, and $6'_m$
symmetries.
Arrows indicate the directions of the surface-normal magnetizations
relative to a global Cartesian frame.
A hinge current flowing into or out of the page
is shown with a cross or dot respectively.
(a) Triangular
prism construction consistent with $3^{\prime}_m$ symmetry.
Surfaces have no magnetization. (b) Square prism
construction consistent with $4^{\prime}_m$ symmetry.
Surface magnetizations form a two-in two-out configuration with
$M_{\perp}=I/2$. (c) Hexagonal prism construction consistent
with $6^{\prime}_m$ symmetry. Surface magnetizations
form a three-in three-out configuration with $M_{\perp}=I/2$.}
\label{fig:screws}
\end{figure}

For these operations, we construct crystallites in the shape of
regular triangular, square, or hexagonal
prisms, respectively,
as shown schematically in Fig.~\ref{fig:screws}.
Each prism is infinite along $z$ and is constructed
to respect the corresponding screw symmetry applied to the rod as a whole.
In the case of $3^{\prime}_m$, symmetry dictates that each facet have
vanishing magnetization. For the $4^{\prime}_m$ and $6^{\prime}_m$
screws, the surface magnetizations may be found from the currents on the
prism hinges.

\paragraph{Symmetries of type $S^+_{\hat{\textbf{z}}}$.}

We construct a surface as in the $S^-_{\hat{\textbf{z}}}$ case. Although $S^+_{\hat{\textbf{z}}}$ operations
leave the surface invariant, they do not flip $M_{\perp}$. We
therefore cannot extract any meaningful
information from this construction.

\begin{figure}[t]
\centering
\includegraphics[width=\columnwidth]{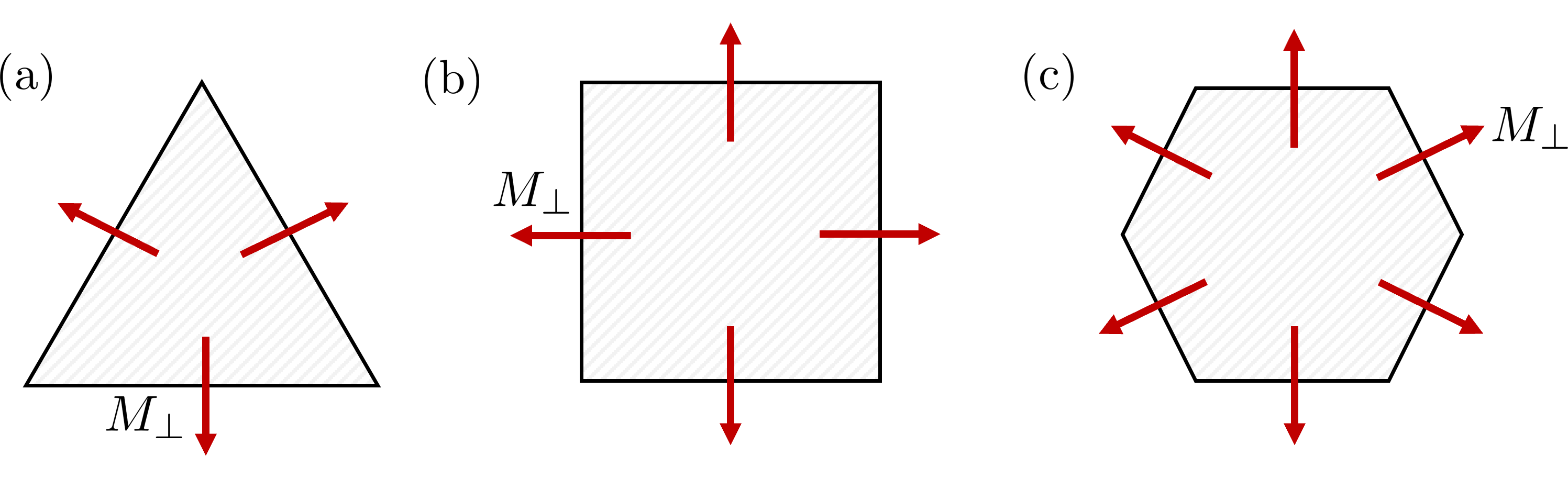}
\caption{
Symmetry constraints arising from $3_m$, $4_m$,
and $6_m$ symmetries. Conventions are the same
as in Fig.~\ref{fig:screws}.
(a) Triangular prism constructed using $3_m$ symmetry.
(b) Square prism constructed using $4_m$ symmetry.
(c) Hexagonal prism constructed using $6_m$ symmetry.
In each case all facets have the same surface magnetization $M_\perp$,
whose magnitude cannot be determined since hinge currents are
absent.
}
\label{fig:antiscrews}
\end{figure}

However, operations in this class may also relate surfaces with normals
not along $\hat{\textbf{z}}$.  For the case of $C_n$ operations with $n=3, 4, 6$, we
may generate the same infinite prism constructions as in the
$N^-_{\hat{\textbf{z}}}$ class of operations. However, as we see in
Fig.~\ref{fig:antiscrews}, all the hinge currents vanish, and although
we know the relative orientations of the facet magnetizations, we cannot
extract any information on their magnitudes as a result.

\begin{figure}[b]
\centering
\includegraphics[scale=0.5]{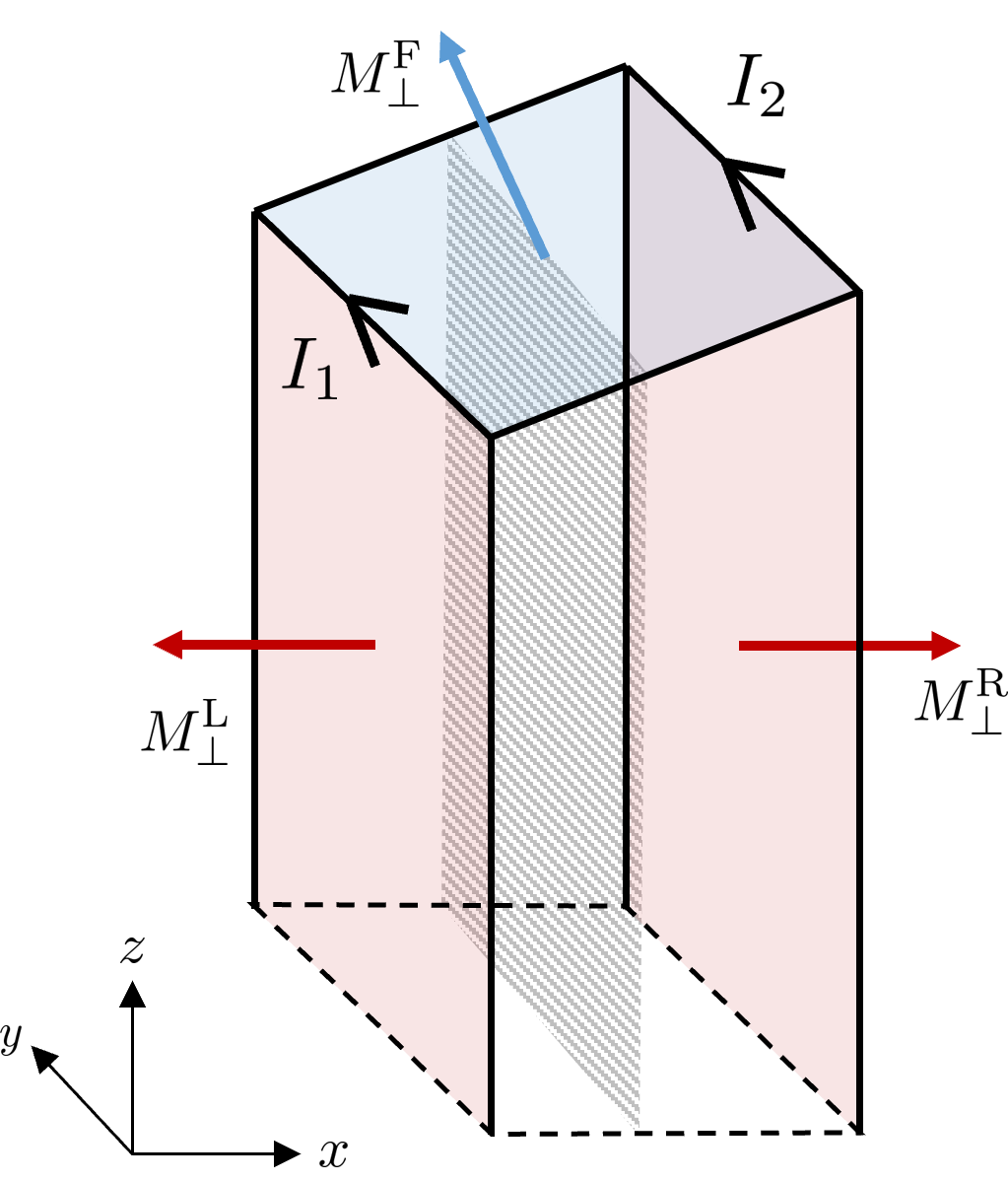}
\caption{Schematic of a slab geometry respecting
an $m'_x$ symmetry, with the mirror plane illustrated by the thatched gray plane bisecting the slab. An additional facet of arbitrary orientation appears at the top. The $m'_x$ symmetry ensures that the magnetizations on the left and right
facets are equal and antiparallel. In this case the knowledge of the hinge currents is insufficient to determine the $M_\perp$ values, as discussed in the text.}
\label{fig:antislabs}
\end{figure}

For the $C_2$ as well as $m^{\prime}_d$ operations, we may construct a slab geometry as in
the case of $R^-_{\hat{\textbf{z}}}$ operations, except that the
$z$-direction will be periodic, and the slab is consistent with either the $C_2$ or
the $m^{\prime}_d$ symmetry. As a result, any possible surface
magnetizations will be equal and antiparallel on the two surfaces, as shown
for the particular case of an $m'_x$ symmetry in Fig.~\ref{fig:antislabs}. With the introduction of an additional
arbitrary surface facet, we may attempt to solve a system of equations

\begin{align}
    0&=M^{\text{L}}_{\perp}-M^{\text{R}}_{\perp},\nonumber\\ I_1&=M^{\text{L}}_{\perp}-M^{\text{F}}_{\perp},\\
    I_2&=M^{\text{F}}_{\perp}-M^{\text{R}}_{\perp},\nonumber
\end{align}
involving the hinge currents. However, this system is unsolvable.

Thus, $S^+_{\hat{\textbf{z}}}$ operations alone are unable to unambiguously define surface magnetizations.

\paragraph{Symmetries of type $N^+_{\hat{\textbf{z}}}$.}

Being nonsymmorphic operations in the $\hat{\textbf{z}}$ direction, no
surface with the same unit normal will ever be preserved by an
$N^+_{\hat{\textbf{z}}}$ operation, so these operations by themselves
tell us nothing about magnetizations at this surface. Furthermore, they
provide no information on surfaces transverse to the $\hat{\textbf{z}}$
surface, as the arguments in Figs.~\ref{fig:antiscrews}
and~\ref{fig:antislabs} apply.

\paragraph{Symmetries of type $R^+_{\hat{\textbf{z}}}$.}

The same arguments as presented for the $S^+_{\hat{\textbf{z}}}$ and 
$N^+_{\hat{\textbf{z}}}$ cases hold here as well. We therefore cannot
define surface magnetizations in this case either.

\subsection{Discussion}
\label{sec:discussion}

Within a classical context, our symmetry analysis reveals that we may unambiguously define magnetizations on the surfaces only when the bulk symmetry group features $S^{-}_{\hat{\textbf{n}}}$, $R^{-}_{\hat{\textbf{n}}}$, or $N^{-}_{\hat{\textbf{n}}}$ operations. 

In these cases, the argument proceeded via a thought experiment
involving the construction of a slab or rod in such a way that the surface facets are related to each other by a global application of the relevant
bulk symmetry. The facet magnetizations could then be determined from a knowledge of the hinge currents in these geometries. We emphasize that, for these bulk symmetries,
the surface magnetizations are still well defined even for 
configurations in which different facets are terminated in 
different ways.
The conclusion that $M_\perp$ is well defined follows just from knowing
that it is possible, in principle, to prepare a globally symmetric
slab or rod.

As it happens, the $S^{-}_{\hat{\textbf{n}}}$,
$R^{-}_{\hat{\textbf{n}}}$, and $N^{-}_{\hat{\textbf{n}}}$ symmetries that allow an unambiguous definition
are precisely the ones that are classified as axion-odd according to Table~\ref{tab:SymmLabels}.
In other words, these are the symmetries that reverse the sign
of the Chern-Simons axion (CSA) coupling,
which describes a topological contribution to the isotropic
linear magnetoelectric coupling of the
material \cite{qi-prb2008,essin-prl2009,varnava-prb2020}.
This is characterized by an ``axion angle'' $\theta_{\text{CS}}$ that
takes the form
\begin{equation}
\theta_{\text{CS}}=-\frac{1}{4\pi}\int_{\text{BZ}}\epsilon_{\mu\nu\sigma}\text{Tr}\left[A_{\mu}\partial_{\nu}A_{\sigma}-\frac{2i}{3}A_{\mu}A_{\nu}A_{\sigma}\right]d\textbf{k},
\end{equation}
where the integration is performed over the Brillouin zone and
$A^{mn}_{\mu}(\textbf{k})=i\langle
u_{m\textbf{k}}|\partial_{\mu}u_{n\textbf{k}}\rangle$ is the Berry
connection matrix, with $|u_{n\textbf{k}}\rangle$ being the
cell-periodic part of the Bloch function of occupied band $n$.

A defining feature of the CSA coupling is that it is gauge-invariant
modulo $2\pi$ under a unitary mixing of the occupied ground state Bloch
functions. That is, if we perform the transformation \begin{equation}
|\tilde{u}_{n\textbf{k}}\rangle=\sum^{\text{occ}}_mU_{mn}(\textbf{k})|u_{m\textbf{k}}\rangle,
\end{equation} where $U_{mn}(\textbf{k})$ is a unitary matrix, then
$\theta_{\text{CS}}\to\theta_{\text{CS}}+2\pi n$ where $n$ is an
integer. The operations of the type $S^{-}_{\hat{\textbf{n}}}$,
$R^{-}_{\hat{\textbf{n}}}$, and $N^{-}_{\hat{\textbf{n}}}$ reverse the sign of
$\theta_{\text{CS}}$; but since the coupling is defined
modulo $2\pi$, it follows that $\theta_{\text{CS}}=0\text{ or }\pi$. Thus, the axion  angle is quantized to values of 0 or $\pi$ in the presence of axion-odd symmetries. By contrast, an axion-even symmetry (simple rotation or  time-reversed improper rotation) allows for a generically nonzero value of $\theta_{\text{CS}}$.

This identification of an intimate connection between
the ability to define unambiguous surface magnetizations and the
presence of a quantized axion coupling is one of the principle
contributions of this work.

While we have been focusing on the consequences of
symmetry in this section, this can only take us so far.
Even in the axion-odd case where we expect a unique definition of
surface magnetization, the symmetry arguments and classical theory do not answer the question of how, in principle, this quantity can be obtained from a direct calculation using only information about the surface Hamiltonian and surface electronic structure of a given surface.
In the opposite case that no axion-odd symmetry is present,
the same question arises, but now there is the additional issue of a
possible shift freedom in the definition of $M_\perp$.
To go beyond the symmetry arguments, we need to develop an explicit
framework for carrying out calculations of surface orbital magnetization
in the general case.  The remainder of this work is devoted to an
exploration of the use of quantum-mechanical orbital-magnetization local markers for the construction of such a framework.

\section{Methods}\label{Methods}

We now turn to a discussion of the computational aspects of surface
magnetization and hinge currents. We expand on the local marker
formulation of orbital magnetization \cite{bianco-prl2013}
and discuss its application in
computing surface magnetization. The section is concluded with a
discussion of microscopic currents and an explanation of their
relationship to the macroscopic hinge current.

\subsection{Local-marker formulation}\label{Markers}

We now work in 3D, where Eq.~(\ref{eq:MarkerEq}) takes the form
\begin{equation}
\textbf{M}=\frac{1}{V}\int_V\mathbfcal{M}(\textbf{r})d\textbf{r}.\label{eq:MarkerEq3d}
\end{equation}
Here $V$ is the unit cell volume and both $\textbf{M}$ and
$\mathbfcal{M}(\textbf{r})$ have units of magnetic moment per unit volume.
(Quantities denoted as $M_\perp$ or $M^\textrm{surf}$
will continue to
be 2D surface magnetizations with units of magnetic moment per unit
area.)
We focus initially on the single component $M_z$ of
$\textbf{M}$,
dropping the $z$ subscript for conciseness.
The local-marker formulation is a direct result of expressing the orbital
magnetization in terms of the single particle density matrix
$P(\textbf{r},\textbf{r}')$.
Bianco and Resta~\cite{bianco-prl2013}
demonstrated that for bulk insulating systems at $T=0$ at the
single-particle level, where $P^2=P$, the magnetization $M$
may be written as

\begin{equation}
M=M_{\text{LC}}+M_{\text{IC}}+M_{\text{C}}
\label{eq:RestaFormula}
\end{equation}
with
\begin{eqnarray}
M_{\text{LC}}&=&-\frac{e}{\hbar V}\text{Im}\text{Tr}[PxQHQyP], \label{eq:LCint} \\
M_{\text{IC}}&=&\frac{e}{\hbar V}\text{Im}\text{Tr}[QxPHPyQ], \label{eq:ICint} \\
M_{\text{C}} &=&-\frac{2\mu e}{\hbar V}\text{Im}\text{Tr}[QxPyQ], \label{eq:Cint}
\end{eqnarray}
where $e<0$ is the electron charge,\footnote{The convention for the elementary charge in the paper by Bianco and Resta \cite{bianco-prl2013} is that $e>0$. As a result, our expressions in Eqs.~(\ref{eq:LCint}-\ref{eq:Cint}) differ by a sign from those introduced in Ref. \cite{bianco-prl2013}.}
the trace is taken per unit cell of volume $V$,
$Q=1-P$, $H$ is the Hamiltonian,
and $\mu$ is the Fermi energy.
$M_{\text{LC}}$, $M_{\text{IC}}$,
and $M_{\text{C}}$ correspond to the so-called local circulation,
itinerant circulation, and Chern number contributions to the
magnetization, respectively \cite{ceresoli-prb2006}.

Eq.~(\ref{eq:MarkerEq3d}), expressing the total
$M$ as a trace of a local marker $\mathcal{M}(\textbf{r})$
over position space, is evidently satisfied by choosing
\begin{equation}
\mathcal{M}(\textbf{r})=\mathcal{M}_{\text{LC}}(\textbf{r})+\mathcal{M}_{\text{IC}}(\textbf{r})+\mathcal{M}_{\text{C}}(\textbf{r}),
\label{eq:Mrsum}
\end{equation}
with
\begin{eqnarray}
\mathcal{M}_{\text{LC}}(\textbf{r})&=&-\frac{e}{\hbar}\text{Im}\langle\textbf{r}|PxQHQyP|\textbf{r}\rangle,\label{eq:LC}\\
\mathcal{M}_{\text{IC}}(\textbf{r})&=&\frac{e}{\hbar}\text{Im}\langle\textbf{r}|QxPHPyQ|\textbf{r}\rangle,\label{eq:IC}\\
\mathcal{M}_{\text{C}}(\textbf{r})&=&-\frac{2\mu
e}{\hbar}\text{Im}\langle\textbf{r}|QxPyQ|\textbf{r}\rangle. \label{eq:Chern}
\end{eqnarray}
Note that
$\mathcal{M}_{\text{C}}(\textbf{r})$ is directly proportional to the
local Chern marker $C(\textbf{r})$ \cite{bianco-prb2011,rauch-prb2018}, i.e.,
\begin{equation}\mathcal{M}_{\text{C}}(\textbf{r})=-\frac{2\mu
e}{\hbar}\text{Im}\langle\textbf{r}|QxPyQ|\textbf{r}\rangle=-\frac{e\mu}{h}C(\textbf{r}),\label{eq:ChernMarker}\end{equation}
consistent with the discussion towards the end of Sec.~\ref{sec:discussion}.

It is important to note that due to the invariance of the trace under
cyclic permutations of operators, different expressions for
$\mathcal{M}(\textbf{r})$ stemming from Eq.~(\ref{eq:RestaFormula}) are
possible. For example, $\mathcal{M}_{\text{LC}}(\textbf{r})$
in Eq.~(\ref{eq:Mrsum})  could be replaced by

\begin{equation}
\mathcal{M}^{\cal L}_{\text{LC}}(\textbf{r})=
-\frac{e}{\hbar}\text{Im}\langle\textbf{r}|HQyPxQ|\textbf{r}\rangle
\label{eq:LCLf} 
\end{equation}
or
\begin{equation}
\mathcal{M}^{\cal R}_{\text{LC}}(\textbf{r})=
-\frac{e}{\hbar}\text{Im}\langle\textbf{r}|QyPxQH|\textbf{r}\rangle,
\label{eq:LCRf} 
\end{equation}
achieved by
applying a cyclic permutation to $\mathcal{M}_{\text{LC}}(\textbf{r})$
in Eq.~(\ref{eq:LC}) and using the fact that $Q^2=Q$.
The $\cal L$ and $\cal R$ superscripts indicate that $H$ has
been moved to the ``left'' or ``right'' respectively, and henceforth we shall
refer to the marker in \eq{LC} as $\mathcal{M}^{\cal C}_{\text{LC}}$
($\cal C$ for ``center'').
Integrating \eqo{LCLf}{LCRf} over all space yields \eq{LCint}, just
as integrating \eq{LC} does.
We refer to the different expressions for the local marker as
trace projections.  Such an ambiguity in the local marker may be
viewed as a manifestation of the fact that microscopic magnetic
dipole densities are not well defined.

As we will see in the following subsections, there is no such
ambiguity with respect to cyclic permutation of operators for the local Chern marker $C(\rr)$, which is therefore
well defined at a specific $\rr$ in the context of a marker-based
theory.  As a result of Eq.~(\ref{eq:ChernMarker}), this implies
that $\mathcal{M}_{\text{C}}(\rr)$ has a linear dependence on the
chemical potential $\mu$ as it is scanned across the gap.  This
observation allowed Zhu et al.~\cite{zhu-prb2021} to introduce a
uniquely defined ``surface magnetic compressibility''
$dM_\perp/d\mu$ in terms of the presence of a net coarse-grained
Chern-marker concentration at the surface.
We note in passing that the Chern marker is closely related to
the local anomalous Hall conductivity, i.e., the
antisymmetric part of the surface conductivity tensor
$\sigma_{ij}(\rr)=\partial j_i(\rr)/\partial E_j$ describing the
first-order response of the local current density ${\bf j}(\rr)$
to a homogeneous macroscopic electric field $\bf E$.  However, as
shown by Rauch et al.~\cite{rauch-prb2018}, the two are not identical,
since the local anomalous Hall conductivity also contains a non-geometric or
``cross-gap'' term that is not captured by the Chern marker.

In the following subsections, we discuss several physical
requirements that we impose on the markers, allowing us to
reduce the number of candidates to just a few that can be
regarded as physically acceptable.

\subsubsection{Independence of origin}
\label{orig}
We first require
that the trace projections should be independent of origin.
A potential marker $\mathcal{M}^{[P,Q,H]}(\rr)$ may be regarded
as depending on the system-specific operators $P$, $Q$,
and $H$ as indicated by the superscript notation.  Introducing the
unitary operator $T$ for a translation by displacement $\bf t$,
and defining translated operators $\tilde{P}=TPT^\dagger$ and
similarly for $Q$ and $H$, the desired independence of origin
is equivalent to asking that $\mathcal{M}^{[P,Q,H]}(\rr)=
\mathcal{M}^{[\tilde{P},\tilde{Q},\tilde{H}]}(\rr+{\bf t})$.
We accomplish this by insisting that the markers be written
in terms of the operators
\begin{equation}
X=PxQ,\quad Y=PyQ,\quad Z=PzQ\label{eq:XYZ}
\end{equation}
and their Hermitian conjugates. A combination like $P\rr Q$ transforms to
$\tilde{P}\tilde{\rr}\tilde{Q}=\tilde{P}(\rr-{\bf t})\tilde{Q}
=\tilde{P}\rr\tilde{Q}$, where the last equality follows from
the fact that $PQ=0$.  Thus, expressions built out of the operators
in Eq.~(\ref{eq:XYZ}) will automatically satisfy the desired
independence of origin.
With this notation,
Eqs.~(\ref{eq:LC}), (\ref{eq:LCLf}), and (\ref{eq:LCRf}) become
\begin{eqnarray}
\mathcal{M}^{\cal C}_{\text{LC}}(\rr)&=&
  -\frac{e}{\hbar}\text{Im}\me{\rr}{XHY^\dagger}{\rr}, \label{eq:LCCo}\\
\mathcal{M}^{\cal L}_{\text{LC}}(\rr)&=&
  -\frac{e}{\hbar}\text{Im}\me{\rr}{HY^\dagger X}{\rr}, \label{eq:LCL}\\
\mathcal{M}^{\cal R}_{\text{LC}}(\rr)&=&
  -\frac{e}{\hbar}\text{Im}\me{\rr}{Y^\dagger XH}{\rr}. \label{eq:LCR}
\end{eqnarray}
Other choices such as $\me{\rr}{yPxQHQ}{\rr}$
would not satisfy this property.
At this point, then, we have three candidates for the LC marker,
and there is a corresponding set of three choices for the IC marker.

As for the Chern-marker contribution, similar considerations imply that
$\mathcal{M}_{\text{C}}(\textbf{r})$ should also be written
in terms of $X$ and $Y$ operators.  Eq.~(\ref{eq:Chern}) then leads to
$\mathcal{M}_{\text{C}}(\textbf{r})=-(2\mu e/\hbar) \text{Im}\langle\textbf{r}|X^\dagger Y|\textbf{r}\rangle
=(2\mu e/\hbar) \text{Im}\langle\textbf{r}|Y^\dagger X|\textbf{r}\rangle$.
Further algebra demonstrates that these expressions are identical with
$\mathcal{M}_{\text{C}}(\textbf{r})=(2\mu e/\hbar) \text{Im}\langle\textbf{r}|X Y^\dagger|\textbf{r}\rangle
=-(2\mu e/\hbar) \text{Im}\langle\textbf{r}|Y X^\dagger|\textbf{r}\rangle$ (see Eq. (A14) of the Appendix of Ref. \cite{rauch-prb2018}).
Thus, we are left with a unique expression for $\mathcal{M}_{\text{C}}(\textbf{r})$, as well as the local Chern marker itself.

\subsubsection{Covariance under rotations}
\label{rotat}

We next insist that the markers should transform as vectors
under global rotations of the system. That is, given a spatial rotation $\mathcal{R}$ and its corresponding unitary operator $\mathfrak{R}$, we ask that $\mathcal{M}^{[P,Q,H]}(\rr)=
\mathcal{R}^{-1}\mathcal{M}^{[\tilde{P},\tilde{Q},\tilde{H}]}
(\mathcal{R}\rr)$, where $\tilde{P}=\mathfrak{R}P\mathfrak{R}^\dagger$, etc.  In particular, we insist that
any candidate for the marker $z$ component should be invariant
under rotation by an arbitrary angle $\theta$ about $\hat{\bf z}$.
For $\theta=\pi/2$
we have $X\rightarrow Y$ and $Y\rightarrow-X$, and using the general formula
$\text{Im}\ev{\cal O}=-\text{Im}\ev{\cal O^\dagger}$, we find that
$\mathcal{M}^{\cal C}_{\text{LC}}$ is invariant while
$\mathcal{M}^{\cal L}_{\text{LC}}$ and
$\mathcal{M}^{\cal R}_{\text{LC}}$ transform into one another.
Thus, neither $\mathcal{M}^{\cal L}_{\text{LC}}$ nor
$\mathcal{M}^{\cal R}_{\text{LC}}$ satisfies rotational invariance
by itself, but at least for $\theta=\pi/2$, the average
$(\mathcal{M}^{\cal L}_{\text{LC}}+\mathcal{M}^{\cal R}_{\text{LC}})/2$
does.

A more careful check confirms that this combination
is also invariant under arbitrary rotations by angle $\theta$.  We can
therefore propose two valid local-circulation markers,
\begin{eqnarray}
\mathcal{M}^{\cal C}_{\text{LC},z}(\rr)&=&
  -\frac{e}{\hbar}\text{Im}\me{\rr}{XHY^\dagger}{\rr}, \label{eq:LCC} \\
\mathcal{M}^{\cal E}_{\text{LC},z}(\rr)&=&
  -\frac{e}{2\hbar}\text{Im}\me{\rr}{\{H,Y^\dagger X\}}{\rr}, \label{eq:LCE}
\end{eqnarray}
where the $z$ subscript has temporarily been restored, and
the $\cal E$ superscript in the second equation involving the
anticommutator indicates that $H$ is in an ``external'' (left or right) position.
More generally, we can define vector markers $\mathbfcal{M}_{\text{LC}}$
whose $x$ and $y$ components have the same form as in \eqs{LCC}{LCE} but
with a cyclic permutation of the Cartesian indices.  Since any rotation
can be written as a product of three consecutive rotations around
Cartesian axes by Euler angles, and since our markers have the
correct rotation properties under each of these separately, it follows
that the overall rotational properties are guaranteed. That is,
the rotation of the marker is consistent with the rotation of
the system.

The same reasoning applies to the itinerant circulation markers
\begin{eqnarray}
\mathcal{M}^{\cal C}_{\text{IC}}(\rr)&=&
  \frac{e}{\hbar}\text{Im}\me{\rr}{X^\dagger HY}{\rr}, \label{eq:ICC} \\
\mathcal{M}^{\cal E}_{\text{IC}}(\rr)&=&
  \frac{e}{2\hbar}\text{Im}\me{\rr}{\{H,YX^\dagger\}}{\rr}, \label{eq:ICE}
\end{eqnarray}
where we have reverted to dropping the $z$ subscript.  Combining with
\eqs{LCC}{LCE}, we can arrive at several suitable expressions for
the full marker in \eq{Mrsum}.  If we consistently use the center
or external versions, we arrive at two choices
\begin{eqnarray}
\mathcal{M}^{\mathcal{C},\mathcal{C}}(\textbf{r}) &=&
  \mathcal{M}^{\mathcal{C}}_{\text{LC}}(\textbf{r})
  +\mathcal{M}^{\mathcal{C}}_{\text{IC}}(\textbf{r}), \label{eq:CCmark} \\
\mathcal{M}^{\mathcal{E},\mathcal{E}}(\textbf{r}) &=&
  \mathcal{M}^{\mathcal{E}}_{\text{LC}}(\textbf{r})
  +\mathcal{M}^{\mathcal{E}}_{\text{IC}}(\textbf{r}). \label{eq:EEmark}
\end{eqnarray}
We also consider a third marker involving a mixed choice,
\begin{eqnarray}
\mathcal{M}^{\mathcal{C},\mathcal{E}}(\textbf{r}) &=&
  \mathcal{M}^{\mathcal{C}}_{\text{LC}}(\textbf{r})
  +\mathcal{M}^{\mathcal{E}}_{\text{IC}}(\textbf{r}). \label{eq:CEmark}
\end{eqnarray}
We could also introduce its partner $\mathcal{M}^{\mathcal{E},\mathcal{C}}$,
but there are actually only three linearly independent markers,
since it is easy to show that
$\mathcal{M}^{\mathcal{C},\mathcal{E}}
 +\mathcal{M}^{\mathcal{E},\mathcal{C}}
 =\mathcal{M}^{\mathcal{C},\mathcal{C}}
 +\mathcal{M}^{\mathcal{E},\mathcal{E}}$.

For the Chern marker contribution of Eqs.~(\ref{eq:Chern}-\ref{eq:ChernMarker}), it is easy to see that for arbitrary rotations $\theta$ about $\hat{\textbf{z}}$ we have $\mathcal{M}_{\text{C}}(\textbf{r})=-(2\mu e/\hbar) \text{Im}\langle\textbf{r}|X^\dagger Y|\textbf{r}\rangle
\to(2\mu e/\hbar) \text{Im}\langle\textbf{r}|Y^\dagger X|\textbf{r}\rangle=\mathcal{M}_{\text{C}}(\textbf{r})$. We may then define a vector marker $\mathbfcal{M}_{\text{C}}$ analogously to $\mathbfcal{M}_{\text{LC}}$, and the former is found to exhibit rotational covariance for the same reasons as the latter.

\subsubsection{Transformation under $H\rightarrow-H$}
\label{minusH}
Our third requirement is based on the following observation regarding the behavior of microscopic currents. As will be shown later in this section, the microscopic current density at $\textbf{r}$ is given by \beq \textbf{j}(\textbf{r})=\frac{e}{\hbar}\int d\textbf{r}'(\textbf{r}-\textbf{r}')\text{Im}[\langle\textbf{r}'|P|\textbf{r}\rangle\langle\textbf{r}|H|\textbf{r}'\rangle].\label{eq:TotBondCurr}\eeq 
Under a transformation that acts to change the overall sign of $H$, the occupied and unoccupied state manifolds are switched; that is, $P'=Q$ and $Q'=P$. Substituting $H'=-H$ and $P'$ into Eq.~(\ref{eq:TotBondCurr}), and using the facts that $Q=1-P$ and that the expectation value of a Hermitian operator is real, it is easy to see that $\textbf{j}(\textbf{r})$ is left unchanged. Therefore, the surface magnetizations and hinge currents must be left unchanged as well.

When $H\to-H$, we see that $X\to X^{\dagger}$, $Y\to Y^{\dagger}$, and $Z\to Z^{\dagger}$. It is then readily apparent that  $\mathcal{M}^{\mathcal{C},\mathcal{C}}\to\mathcal{M}^{\mathcal{C},\mathcal{C}}$ and $\mathcal{M}^{\mathcal{E},\mathcal{E}}\to\mathcal{M}^{\mathcal{E},\mathcal{E}}$. However, surface magnetization from the $\mathcal{M}^{\mathcal{C},\mathcal{E}}$ marker is not left invariant, as $\mathcal{M}^{\mathcal{C},\mathcal{E}}\to\mathcal{M}^{\mathcal{E},\mathcal{C}}$. We therefore eliminate $\mathcal{M}^{\mathcal{C},\mathcal{E}}$ from the list of valid markers.

We note that for the Chern marker contribution of Eqs.~(\ref{eq:Chern}-\ref{eq:ChernMarker}), under such a transformation $C(\textbf{r})\to -C(\textbf{r})$, while $\mu\to-\mu$. Therefore $\mathcal{M}_{\text{C}}(\textbf{r})$ is left invariant.

\subsubsection{Magnetic quadrupole of finite systems}
\label{MQM_marker}
To conclude our list of acceptable marker properties, we turn to a discussion of the possible role of the bulk magnetic quadrupole moment (MQM) in a theory of surface magnetization. Our motivation for considering the MQM stems from the recently developed theory of boundary electric polarization \cite{benalcazar-prb2017,ren-prb2021}; for a crystallite with vanishing bulk polarization, the theory identifies the bulk electric quadrupole moment as contributing to the electric polarization at the surface. By analogy, it would then be unsurprising if the MQM similarly contributed to the magnetization at a surface for systems with zero bulk magnetization. Earlier works have focused on deriving expressions for
the MQM within periodic boundary conditions \cite{shitade-prb2018,gao-prb2018,gao-prb2018-no2},
but have not discussed whether or how it manifests itself on the
boundary of a bulk. 

As a theory of the bulk MQM density for an extended system
is not well established, we focus here on the case of an ideal
molecular crystal, i.e., a collection of identical independent
units, which we refer to as ``molecules'', arranged without
overlap on a crystal lattice.

Classically, the MQM tensor $\mq_{ij}$ of a single molecule
is defined in terms of its microscopic current distribution
$\textbf{j}(\textbf{r})$ as
\beq
Q_{ij}=\frac{1}{3}\int
(\textbf{r}\times\textbf{j})_i \, r_j\,d\textbf{r}.
\label{eq:MQM}
\eeq
This is a traceless tensor, since

\beq
\mq_{ii}=\frac{1}{3}\int(\textbf{r}\times\textbf{j})
      \cdot\textbf{r}\,d\textbf{r}=0.
\eeq
Its antisymmetric part is known as the toroidal moment, while
its symmetric part
\beq
\widetilde{\mq}_{ij}=\frac{1}{2} (\mq_{ij}+\mq_{ji}),
\eeq
is solely responsible for the quadrupole contribution to the
far $\mathbf{B}$ field in the multipole expansion.

When this molecular unit is periodically repeated to generate a large
but finite crystallite, the quadrupole moment densities are defined
as $\mqdb=\mqb/V$ and $\widetilde{\mqdb}=\widetilde{\mqb}/V$,
where $V$ is the unit cell volume.  One can then show that the MQM
generates a magnetization at a surface with outward normal
$\hat{\textbf{n}}$ given by \cite{raab-book2005}
\beq
M^\textrm{surf}_i=\mqd_{ij}n_j.
\label{eq:MQM_surfMag}
\eeq

Repackaging the toroidal part of the MQM density as the vector
$\mathfrak{m}_i=\epsilon_{ijk}\mqd_{jk}/2$,
it is clear that this produces a surface magnetization
$\mathbf{M}^\textrm{surf}_\parallel=\hat{\textbf{n}}\times\mathfrak{m}$ that
is parallel to the surface. This part does not contribute
to the surface-normal component $M_\perp$, which is just
given by
\beq
M_\perp^{(\hat{\mathbf{n}})} = \mqd_{ij} n_in_j
       = \hat{\mathbf{n}}^\mathrm{T}\cdot\mqdb\cdot\hat{\mathbf{n}} .
\label{eq:m-to-surf}
\eeq
We are thus free to insert either the unsymmetrized $\mqdb$ or the
symmetrized $\widetilde{\mqdb}$ tensor into \eq{m-to-surf} when computing
$M_\perp$ values, ignoring the toroidal component \cite{dubovik-jetp1986,dubovik-pr1990}.

We  will refer to the MQM tensors defined as in \eq{MQM}
as \textit{current-based} quadrupoles.
In the special case that the current distribution takes the form
of a collection of point magnetic dipoles $\mathbf{m}_\mu$ located
at sites $\mathbf{r}_\mu$, one can introduce an alternative
\textit{dipole-based} definition

\beq
Q^\text{dip}_{ij} = \sum_\mu m_{\mu,i} \, r_{\mu,j} \label{eq:Q_dip}.
\eeq
If one considers the limit that the point dipoles are
constructed from small current loops of vanishing size,
it turns out that the traceless part of $Q^\text{dip}$ is identical
with $Q$ (see Appendix~\ref{AppC}).  That is,
\beq
Q_{ij}=Q^\text{dip}_{ij}-\frac{1}{3}\delta_{ij}Q_{kk}^\text{dip} .
\label{eq:dip-nodip}
\eeq

For a molecular crystal,
the dipole formula provides a natural connection to the use of
the local magnetization marker. That is, 
we can calculate the MQM density of the crystal as
\beq
\mqd_{ij}^\text{dip} = \frac{1}{V} \int {\cal M}_i(\mathbf{r}) \, r_j \, d^3r
\label{eq:quad-cont}
\eeq
or
\beq
\mqd_{ij}^\text{dip} = \frac{1}{V} \sum_\mu {\cal M}_{\mu,i} \, r_{\mu,j} ,
\label{eq:quad-tb}
\eeq
where ${\cal M}_i(\mathbf{r})$ and ${\cal M}_{\mu,i}$ are the
continuum and discrete markers, respectively.  Therefore, in
the rest of this paper, we will focus on the dipole-based
expression in \eq{quad-tb} in the context of our tight-binding
calculation of the local markers.  We will then remove the
trace part in order to compare with the current-based
tensor computed from \eq{MQM} in the same TB framework.

Tests of this kind will be presented in Sec.~\ref{MolecQuadMom}.
Unlike the other criteria described above in
Secs.~\ref{orig} to \ref{minusH}, we do not know how to determine
\textit{a priori} whether a certain marker will satisfy the
present criterion of reproducing the MQM correctly for
molecular crystals.  However, we will see in Sec.~\ref{MolecQuadMom}
that some choices of marker
sometimes fail this
test.  Thus, even if numerical in nature, these tests provide
important constraints on the appropriate choice of marker for
the computation of surface magnetization.

We note in passing that the use of the raw dipole-based MQM,
without removing the trace part, will generate different results
for the surface magnetizations $M_\perp$.  However, it will shift
$M_\perp$ equally on all facets, and therefore will still
yield a correct prediction of the hinge currents, as these only
depend on differences of $M_\perp$ on neighboring facets (see Eq.~(\ref{eq:MainHingeEq})).
In this sense, the use of the raw dipole-based MQM in place of the traceless
one is reminiscent of the shift freedom in the definition of
$M_\perp$ discussed in the Introduction,
here for the case of an idealized molecular crystal.

\subsubsection{Discussion}

The considerations of Secs.~\ref{orig}-\ref{minusH} have reduced the list of acceptable markers to the two given in Eqs.~(\ref{eq:CCmark}-\ref{eq:EEmark}),
augmented by the uniquely defined
Chern-marker contribution $\mathcal{M}_{\text{C}}(\rr)$.
The marker $\mathcal{M}^{\mathcal{C},\mathcal{C}}$ of \eq{CCmark}
is the original one of Bianco and Resta~\cite{bianco-prl2013},
while $\mathcal{M}^{\mathcal{E},\mathcal{E}}$ of \eq{EEmark} is a new candidate that has not, to our knowledge, been considered previously.
We also note that there is no restriction on using a linear combination of  $\mathcal{M}^{\mathcal{C},\mathcal{C}}$ and $\mathcal{M}^{\mathcal{E},\mathcal{E}}$ to write down a new marker, as long the linear combination reproduces the bulk magnetization $M$ of Eq.~(\ref{eq:RestaFormula}). To maintain this requirement, the (real) coefficients $a_{\mathcal{CC}}$ and $a_{\mathcal{EE}}$ of the linear combination must sum to unity. 

We select the particular linear combination $(a_{\mathcal{CC}},a_{\mathcal{EE}})=(1/3,2/3)$ to form the ``linear'' marker
 \begin{equation}
    \mathcal{M}^{\text{lin}}(\textbf{r})=\frac{1}{3}\mathcal{M}^{\mathcal{C},\mathcal{C}}(\textbf{r})+\frac{2}{3}\mathcal{M}^{\mathcal{E},\mathcal{E}}(\textbf{r}). \label{eq:Magicmark}
\end{equation}
This marker is also expressible as \begin{equation}
  \mathcal{M}^{\text{lin}}(\textbf{r})=\frac{1}{3}\mathcal{M}^{\mathcal{C},\mathcal{C}}(\textbf{r})+\frac{1}{3}\mathcal{M}^{\mathcal{L},\mathcal{L}}(\textbf{r})+\frac{1}{3}\mathcal{M}^{\mathcal{R},\mathcal{R}}(\textbf{r}),  
\end{equation} i.e. it is the arithmetic average of the $\mathcal{M}^{\mathcal{C},\mathcal{C}}$, $\mathcal{M}^{\mathcal{L},\mathcal{L}}$, and $\mathcal{M}^{\mathcal{R},\mathcal{R}}$ markers.
Such a choice might heuristically be justified by noting that since
we have no reason to prefer any one of \eqr{LCCo}{LCR}, we choose an equally
weighted average of all three of them.
For the rest of this paper, we will focus on the markers $\mathcal{M}^{\mathcal{C},\mathcal{C}}$ and $\mathcal{M}^{\mathcal{E},\mathcal{E}}$
of Eqs.~(\ref{eq:CCmark}-\ref{eq:EEmark}) and $\mathcal{M}^{\text{lin}}$
of Eq.~(\ref{eq:Magicmark}) when reporting the results of numerical tests.

Having formulated our list of markers, we are faced with several questions regarding their behavior. First, it remains to be seen which, if any, of the markers will correctly predict the hinge current. For each of the two surfaces forming a hinge, identical markers will be used to compute their respective magnetizations, which in turn will be substituted into Eq.~(\ref{eq:MainHingeEq}) to yield the hinge current.
In  such tests, it is of interest to check whether there is a dependence on the presence of axion-odd vs.\ axion-even symmetries.
Recall that at the level of classical electromagnetism, only differences of the magnetizations of surface facets sharing a hinge are generally well defined.
As an exception, when axion-odd symmetries are present, the $M_\perp$ values can be determined from a knowledge of the hinge currents.
We should then like to see whether the quantum marker-based theory is also better behaved when axion-odd symmetries are present, and whether, in the absence of such symmetries, the shift freedom of the classical framework reappears at the quantum level.
This would be signaled by the existence of multiple markers correctly predicting the hinge currents but differing by a constant shift for all facets.

Before attempting to answer these questions, we discuss some technicalities
of the computation of the surface magnetization from the local markers.

\subsection{Surface magnetization and macroscopic averaging}
\label{AvgMark}

Recall that we assume that the system has enough symmetry
to force the macroscopic magnetization to vanish in the bulk.
However, this does not necessarily imply that
$\mathcal{M}(\textbf{r})$ vanishes everywhere, only that its
bulk unit cell average vanishes.  In case $\mathcal{M}(\textbf{r})$
does vanish identically deep in the bulk, then for any given
trace projection, it is straightforward to integrate
$\mathcal{M}(\textbf{r})$ over one surface unit cell, with
the surface-normal integral carried deep enough into bulk and
vacuum, to obtain a corresponding macroscopic surface
magnetization.  If this is not the case, however, a coarse-graining
procedure is required, as described next.

We construct an insulating slab of thickness $L$ with
the outward normals at the top and bottom surfaces being
$\hat{\textbf{z}}$ and $-\hat{\textbf{z}}$ respectively, and with
cell-periodic boundary conditions in $x$ and $y$. The slab features the
maximal symmetry allowed by the bulk. We assume $L>>c$,
where $c$ is the lattice constant in the surface-normal direction. We define a
\say{layer-resolved} local marker by averaging the local marker over an in-plane unit cell of area $A$ at fixed $z$:
\begin{equation}
\overline{\mathcal{M}}(z)=\frac{1}{A}\int_A\mathcal{M}(x,y,z)\,dx\,dy.
\label{eq:AvrgMark}
\end{equation}
Note that the total magnetization of the slab (magnetic moment per unit area)
is the integral of $\overline{\mathcal{M}}(z)$ over the
thickness of the slab.

The macroscopic
surface magnetization is determined from the application of a
smoothening procedure for the layer-resolved marker, since simple sums
of the marker may not be convergent. We employ the
sliding window average approach to compute the surface magnetization;
for details on this method, we refer the reader to Sec. IV.A of Ref.~\cite{ren-prb2021}. In numerical work, the averaging amounts to an
integration of the local marker weighted by a \say{ramp function} that
extends sufficiently deep in the bulk. The ramp-down function is defined
as
\begin{equation}
f_d(u)=\begin{cases}
   1,\text{ if }u<0\\
   1-u/d,\text{if }u\in[0,d]\\
   0,\text{ if }u>d
\end{cases}.
\end{equation}

Letting the bottom surface of the slab be located at $z=0$, we then get
\begin{equation}
M_{\perp}=\int\overline{\mathcal{M}}(z)f_c(z-z_0)\,dz,
\label{eq:Msurf}\end{equation}
where the range of $z$ integration runs from deep in the vacuum to deep in the interior of the crystal.
The result is independent of $z_0$ so long as it is
sufficiently deep in the bulk region of the slab.

The formalism for the calculation of a surface magnetization has been
presented here in the context of a continuum-space treatment.
However, our numerical
tests will be performed on discrete TB models; in this setting,
Eqs.~(\ref{eq:AvrgMark}) and~(\ref{eq:Msurf}) are expressed as discrete
sums over TB sites, where the local markers are now defined. For further
details, we refer the reader to Appendix~\ref{AppA}.

\subsection{Macroscopic hinge current and macroscopic averaging}
\label{AvgCurr}

A direct calculation of the macroscopic hinge
current $I$ is necessary in order to test whether it is correctly
predicted by each of the various trace projections. The calculations will be
accomplished via a macroscopic averaging procedure
analogous to that of Sec.~\ref{AvgMark}, but now applied to the integration
of the microscopic current
density over the appropriate hinge region. In this section and in the rest of the paper, we label vector operators with a hat in order to distinguish them from ordinary vectors. 

The microscopic current
density $\textbf{j}(\textbf{r})$ is obtained from the expectation value of
the microscopic current density operator $\hat{\textbf{j}}(\textbf{r})$.
Given a single-particle Hamiltonian $H$, this is given by
\begin{equation}
\hat{\textbf{j}}(\textbf{r})=\frac{e}{2}(|\textbf{r}\rangle\langle\textbf{r}|\hat{\textbf{v}}+\hat{\textbf{v}}|\textbf{r}\rangle\langle\textbf{r}|)
\end{equation}
where $\hat{\textbf{v}}=(1/i\hbar)[\hat{\textbf{r}},H]$ is the
velocity operator. $\textbf{j}(\textbf{r})$ is then obtained as
$\text{Tr}[P\hat{\textbf{j}}(\textbf{r})]$, i.e., the trace of
$\hat{\textbf{j}}(\textbf{r})$ against the singe-particle density matrix
$P$, so that
\begin{equation}
\textbf{j}(\textbf{r})=\frac{e}{\hbar}\int d\textbf{r}'(\textbf{r}-\textbf{r}')\text{Im}[\langle\textbf{r}'|P|\textbf{r}\rangle\langle\textbf{r}|H|\textbf{r}'\rangle].
\label{eq:ContCurr}
\end{equation}

\begin{figure}[t]
\centering
\includegraphics[width=3.5in]{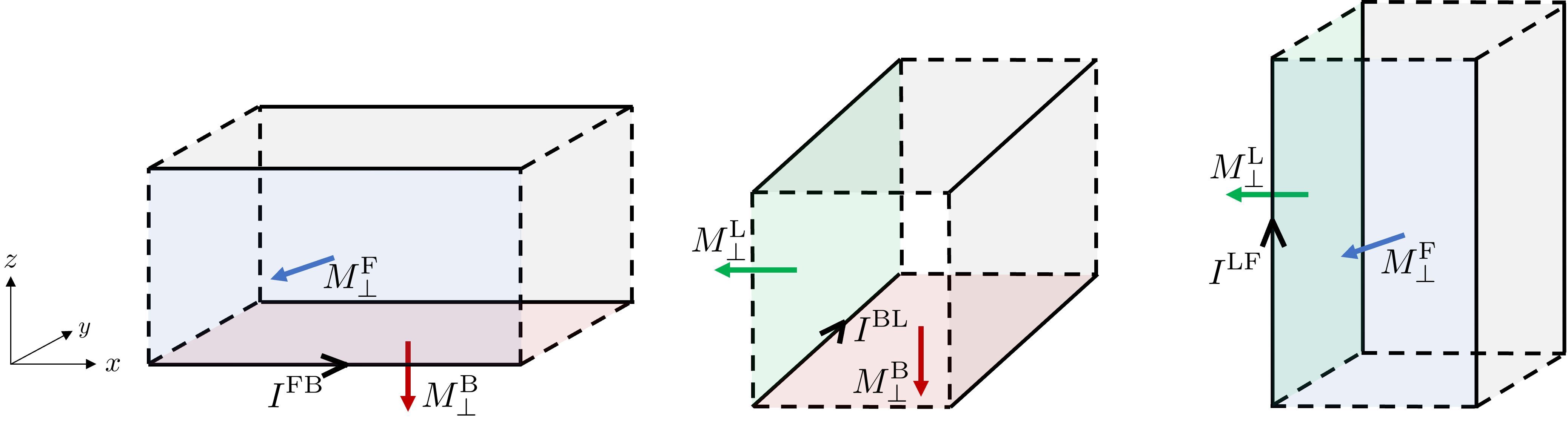}
\caption{Insulating rod geometries utilized in computing macroscopic hinge currents. Each rod is cut out from the bulk, and satisfies periodic boundary conditions along one of the three Cartesian directions. The
surface magnetizations $M^{\text{L}}_{\perp}$,
$M^{\text{B}}_{\perp}$, and $M^{\text{L}}_{\perp}$ are determined from calculations described in Sec.~\ref{AvgMark}. The macroscopic hinge currents $I^{\text{FB}}$, $I^{\text{BL}}$, and $I^{\text{LF}}$ that are computed in the text are indicated by the black arrows.}
\label{fig:Rod_current}
\end{figure}

Our calculations of the macroscopic current are performed on a series of insulating infinite rod geometries that retain periodic boundary conditions along one of the three Cartesian directions, as illustrated in Fig.~\ref{fig:Rod_current}. For concreteness, in this subsection we focus on computing the hinge current $I^{\text{FB}}$ for a rod running parallel to the $x$-axis, with unit-cell periodicity $a$. Let the rod have width $W$ along $y$ and height $L$ along $z$, with $W$ and $L$ much larger than the cell dimensions $b$ and $c$ along these respective dimensions.

In analogy with the surface magnetization, the macroscopic current is
obtained from a sliding window average performed over the microscopic
currents at the corresponding hinge, but now in two
dimensions. The appropriate weight function
$w(y,z)$ is a product of two ramp functions in $y$ and $z$, namely
$w(y,z)=f_b(y-y_0)f_c(z-z_0)$ where $y_0$ and $z_0$ are located
sufficiently deep within the interior of the rod. $I^{\text{FB}}$ is then
given by
\begin{equation}
I^{\text{FB}}=\frac{1}{a}\int^a_0dx\iint dy\,dz\,j_x(x,y,z)w(y,z),\label{eq:MacroCurrIntg}
\end{equation}
where the range of integration is over all $y$ and $z$.

The currents $I^{\text{BL}}$ and $I^{\text{LF}}$ are computed analogously with the appropriate cell parameters and Cartesian variables used in Eq.~(\ref{eq:MacroCurrIntg}). As in the previous subsection, the formalism here is developed in the context of a continuum. In a TB formulation, Eqs.~(\ref{eq:ContCurr}) and (\ref{eq:MacroCurrIntg}) will turn into discrete sums over TB sites. We refer the
reader to Appendix~\ref{AppB} for further details.

\section{Numerical Results}
\label{Results}

To test the results of the
symmetry analysis of Sec.~\ref{Symms} and the local marker calculation
of surface magnetization, we study three tight-binding (TB) models of spinless electrons.
The first model is comprised of an infinite stack of the Haldane-model layers, while the other two
models are composed of stacks of two-dimensional (2D) square-plaquette
layers. Each model features symmetry-enforced zero bulk magnetization and is considered at half-filling. The TB basis orbitals are assumed to have no orbital moment of their own and to diagonalize the position operators.

We focus on the variation of a single parameter in the Hamiltonian away from the special value that quantizes the CSA
coupling, and investigate how the surface magnetization found from the various trace projections behaves as a function of this variation. For each model, the Fermi energy is set to zero, allowing us to neglect the Chern marker
contribution to the local magnetization marker. 

We use the slab and rod geometries outlined in Secs.~\ref{AvgMark} and~\ref{AvgCurr} to compute the surface magnetizations and hinge currents, respectively. Fig.~\ref{fig:Rod_current} illustrates the particular magnetizations and hinge currents that we compute. The surface magnetizations are found from the coarse-graining procedure described in Sec.~\ref{AvgMark} while the macroscopic hinge currents are found from the coarse-graining procedure of Sec~\ref{AvgCurr}. Slab and rod geometries for the computation of
surface magnetizations and hinge currents are generated by truncating
the bulk, i.e., the hoppings to vacant sites are removed while other
hoppings and site energies are unchanged. The electronic Hamiltonians
for bulk, slab, and rod geometries are constructed and solved using the
\verb|PythTB| code package \cite{pythtb}.

Details of our numerical results will be given in Appendix~\ref{AppD}. For each model, however, we provide here two tables to demonstrate the behavior of the surface magnetization and hinge currents for the axion-odd regime, as well as a representative axion-even setup. Each table reports the values of $M^{\text{B}}_{\perp}$, $M^{\text{F}}_{\perp}$, and $M^{\text{L}}_{\perp}$ computed from the different markers. The tables also display the differences $\Delta M^\text{FB}$, $\Delta M^\text{BL}$, and $\Delta M^\text{LF}$ between 2D surface magnetizations at different surface facets found from Eq.~(\ref{eq:Msurf}) using identical markers. For example, $\Delta M^{\text{FB}}$ indicates the difference $M^{\text{F}}_{\perp}-M^{\text{B}}_{\perp}$. These are then compared to the appropriate directly computed hinge currents displayed in the bottom-most rows of the tables. 

At the end of this section, we numerically address the question of which local marker is able to reproduce the current-based MQM for a finite system with no magnetic moment. We perform a test on a finite cubic model, and compare the current-based MQM tensor to the dipole-based MQM tensors derived from the $\mathcal{M}^{\mathcal{C},\mathcal{C}}$, $\mathcal{M}^{\mathcal{E},\mathcal{E}}$, and $\mathcal{M}^{\text{lin}}$ local markers. 

\begin{figure}[t]
\centering
\includegraphics[width=3.5in]{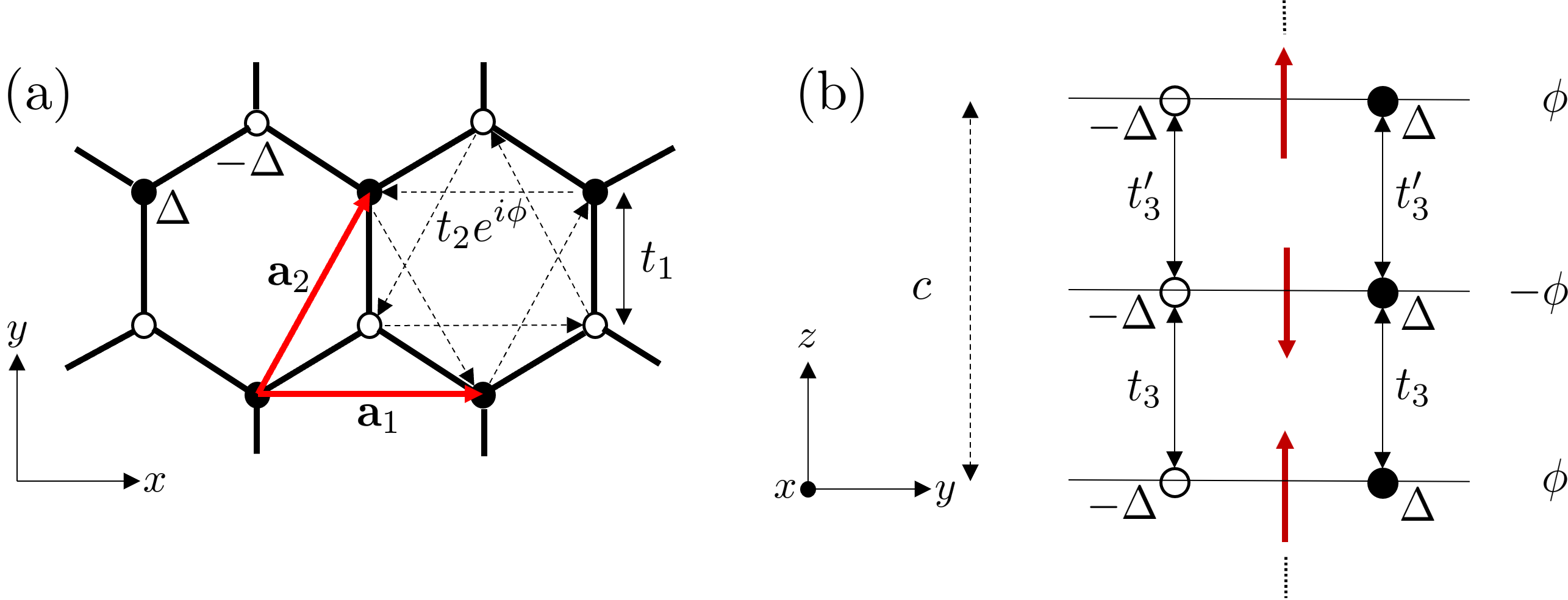}
\caption{(a) Visualization of a single Haldane layer. The different
sublattices are indicated by filled and empty circles, respectively, and
feature onsite potentials $\Delta$ and $-\Delta$. The nearest neighbor
hopping $t_1$ is indicated by the double-sided solid arrow, while the
complex next nearest neighbor hopping $t_2e^{i\phi}$ is indicated by
one-sided dashed arrows. (b) A side view of the bulk unit cell along the
$-x$ direction for the model of Sec~\ref{AltHald}. The vertical
interlayer hoppings connect identical sublattices and are indicated
by solid double-sided arrows that are labeled with the corresponding
hopping amplitude. Crimson arrows indicate the magnetizations of the
layers in the decoupled regime. To the right are the
phases of the complex intralayer hoppings.}
\label{fig:AFM_Hald}
\end{figure}

\begin{table}[b]
\caption{\label{tab:AHL_mags_axodd}Surface magnetizations and hinge currents
(in units of $10^{-5} e/\hbar$) for the alternating Haldane model
with $t'_3=0.1$ (axion-odd).}
\begin{ruledtabular}
\begin{tabular}{lcccccc}
& $M_\perp^\text{B}$ & $M_\perp^\text{F}$ & $M_\perp^\text{L}$ &
$\Delta M^\text{FB}$ & $\Delta M^\text{BL}$ & $\Delta M^\text{LF}$ \\
\hline
$M^{\cal CC}$ & 1.8082 & 0.0000 & 0.0000 & $-$1.8082 & 1.8082 & 0.0000 \\ 
$M^{\cal EE}$ & 1.8082 & 0.0000 & 0.0000 & $-$1.8082 & 1.8082 & 0.0000 \\ 
$M^\text{lin}$ & 1.8082 & 0.0000 & 0.0000 & $-$1.8082 & 1.8082 & 0.0000 \\ 
$I_\text{calc}$ & & & & $-$1.8082 & 1.8082 & 0.0000 
\end{tabular}
\end{ruledtabular}
\end{table}

\subsection{Alternating Haldane model}
\label{AltHald}

The first model we study consist of half-filled Haldane-model layers \cite{haldane-prl1988} placed directly on top of each other, which are then subsequently coupled via interlayer hoppings along the $(001)$ direction. The interlayer hoppings couple sites on identical sublattices, and alternate in value from one interlayer region to the next. The intralayer parameters are chosen to be such that if the layers were decoupled, their Chern numbers would vanish and their orbital magnetizations would form an up-down pattern akin to the magnetic moments of an A-type antiferromagnet. The layers are equidistantly spaced, with two layers per unit cell. Figs.~\ref{fig:AFM_Hald}(a)-(b) provide detailed illustrations of the lattice structure as well as the hoppings and onsite energies of the model. In our calculations we set the lattice vectors as $\textbf{a}_1=\hat{\textbf{x}}$, $\textbf{a}_2=\frac{1}{2}\hat{\textbf{x}}+\frac{\sqrt{3}}{2}\hat{\textbf{y}}$, and $\textbf{a}_3=\hat{\textbf{z}}$.

The symmetries of this model ensure that the bulk magnetization
is zero. In particular, there is a three-fold rotation axis $C_{3z}$
about the lattice sites (or honeycomb centers) as well as a
time-reversed mirror plane $m'_z$ passing between the Haldane layers
that eliminate the magnetization. Additionally, a time-reversed mirror
plane $m'_x$ bisects the honeycombs along their main diagonals, and a
two-fold rotation axis $C_{2y}$ lies between the layers and in the plane
of the $m'_x$ mirror.

The quantization of the CSA coupling is controlled by varying
$t'_3$ while keeping all other parameters fixed.
When $t^{\prime}_3=t_3$, the system
gains a time-reversed half-translation $\{E^{\prime}|c/2\}$ symmetry as well as
an $m_z$ mirror plane residing in the layers; both symmetries are
axion-odd.

We adopt the parameter values $\Delta=-1.5$, $t_1=-0.1$, $t_2=0.15$, $\phi=\pi/4$, $t_3=0.1$, and $0.05\leq t^{\prime}_3\leq0.15$.
$M^{\text{B}}_{\perp}$ and $M^{\text{F}}_{\perp}$ are found from slabs with six unit cells along $\textbf{a}_3$ and eighteen unit cells along $\textbf{a}_2$, respectively. The calculation of $M^{\text{L}}_{\perp}$ involves the construction of a supercell with lattice vectors $\textbf{A}_1=\textbf{a}_1$, $\textbf{A}_2=-\textbf{a}_1+2\textbf{a}_2$, and $\textbf{A}_3=\textbf{a}_3$. A slab with fifteen cells along $\textbf{A}_1$ is then constructed. For $M^{\text{B}}_{\perp}$, $M^{\text{F}}_{\perp}$, and $M^{\text{L}}_{\perp}$, $20\times20$, $15\times15$, and $5\times5$ $\textbf{k}$-space meshes in reduced coordinates are used, respectively.

\begin{table}[b]
\caption{\label{tab:AHL_mags_axeven}Surface magnetizations and hinge currents
(in units of $10^{-5} e/\hbar$) for the alternating Haldane model
with $t'_3=0.15$ (axion-even).}
\begin{ruledtabular}
\begin{tabular}{lcccccc}
& $M_\perp^\text{B}$ & $M_\perp^\text{F}$ & $M_\perp^\text{L}$ &
$\Delta M^\text{FB}$ & $\Delta M^\text{BL}$ & $\Delta M^\text{LF}$ \\
\hline
$M^{\cal CC}$ & 1.7970 & 0.0830 & 0.0163 & $-$1.7140 & 1.7807 & $-$0.0667 \\ 
$M^{\cal EE}$ & 1.7970 & 0.0830 & 0.0163 & $-$1.7140 & 1.7807 & $-$0.0667 \\ 
$M^\text{lin}$ & 1.7970 & 0.0830 & 0.0163 & $-$1.7140 & 1.7807 & $-$0.0667 \\ 
$I_\text{calc}$ & & & & $-$1.7140 & 1.7807 & $-$0.0667
\end{tabular}
\end{ruledtabular}
\end{table}

The macroscopic hinge current $I^{\text{FB}}$ is found from an $x$-extensive rod composed of twenty cells along $\textbf{a}_3$ and twenty cells along $\textbf{a}_2$. The $y$-extensive rod employed for $I^{\text{BL}}$ is composed of twenty cells along both $\textbf{A}_1$ and $\textbf{A}_3$, while the $z$-extensive rod for $I^{\text{LF}}$ is composed of twenty cells each along $\textbf{A}_1$ and $\textbf{A}_2$. For each rod, a 1D $\textbf{k}$-space sampling of thirty points is employed.

For $t'_3=0.1$ and $t'_3=0.15$, respectively, Tables~\ref{tab:AHL_mags_axodd} and~\ref{tab:AHL_mags_axeven} report the values of the surface magnetizations computed from the different markers. From the tables, we note that for a given surface all three markers yield identical magnetizations, independent of whether or not the system features an axion-odd symmetry. Furthermore, the magnetizations always yield the correct values of the hinge currents. In the axion-odd case of $t'_3=0.1$, we observe that $M^{\text{F}}_{\perp}$ and $M^{\text{L}}_{\perp}$ may be understood to be zero due to the slab constructions for their calculations featuring the $m_z$ mirror plane. By the same token, $I^{\text{LF}}$ vanishes as well, as the $z$ extensive rod used for its calculation also features the same mirror symmetry.

\subsection{Two-layer square-plaquette model}
\label{2plaq}

\begin{table}[b]
\caption{\label{tab:2plaq_straight_mags_axodd}Surface magnetizations and hinge currents (in units of $10^{-6} e/\hbar$) for the two-layer plaquette model with straight-edge terminations and with $t'_3=0.1$ (axion-odd).}
\begin{ruledtabular}
\begin{tabular}{lcccccc}
& $M_\perp^\text{B}$ & $M_\perp^\text{F}$ & $M_\perp^\text{L}$ &
$\Delta M^\text{FB}$ & $\Delta M^\text{BL}$ & $\Delta M^\text{LF}$ \\
\hline
$M^{\cal CC}$ & 2.3768 & 0.0000 & 0.0000 & $-$2.3768 & 2.3768 & 0.0000 \\ 
$M^{\cal EE}$ & 2.3768 & 0.0000 & 0.0000 & $-$2.3768 & 2.3768 & 0.0000 \\ 
$M^\text{lin}$ & 2.3768 & 0.0000 & 0.0000 & $-$2.3768 & 2.3768 & 0.0000 \\ 
$I_\text{calc}$ & & & & $-$2.3768 & 2.3768 & 0.0000 
\end{tabular}
\end{ruledtabular}
\end{table}

\begin{figure}[t]
\centering
\includegraphics[width=3.5in]{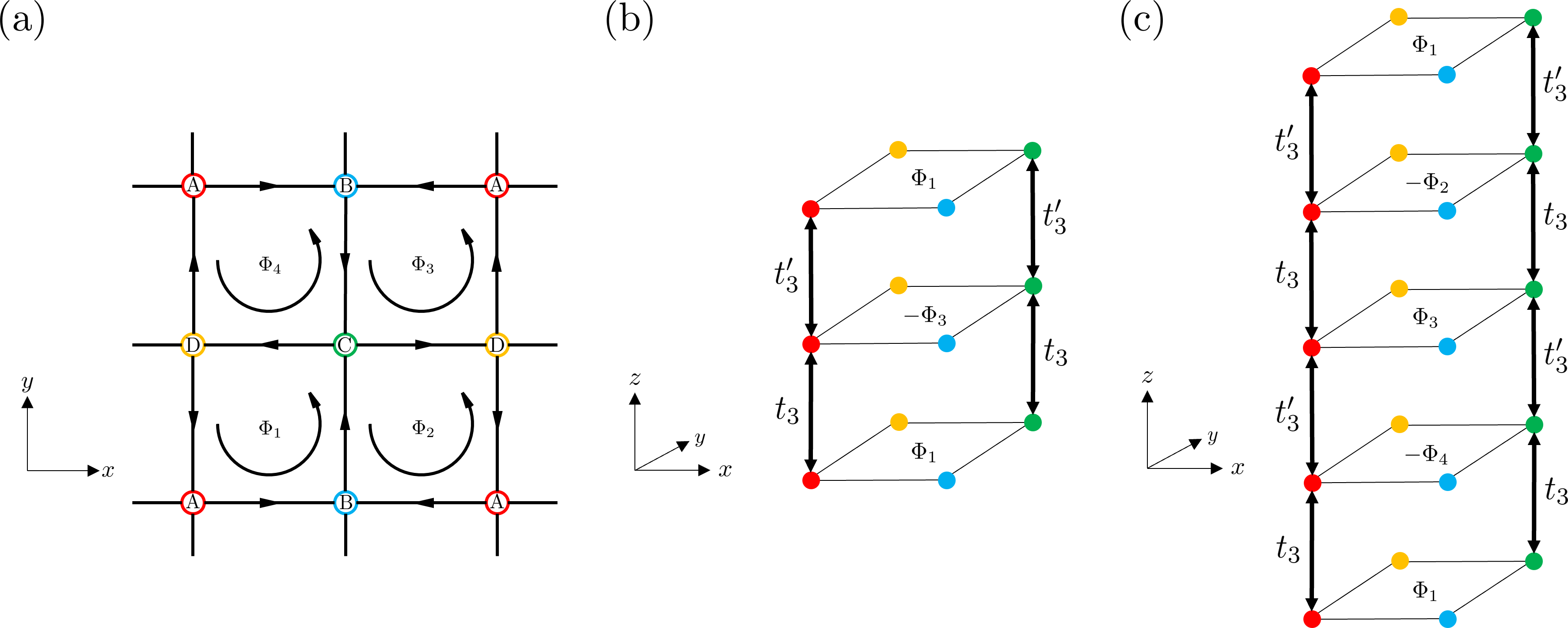}
\caption{(a) Primitive cell of the 2D square plaquette model. The
hoppings between different lattice sites, marked by solid straight lines
with arrows, are such that the total flux through the cell is an integer
multiple of $2\pi$.
(b) Two-layer plaquette model.  Only A-A and C-C interlayer hoppings
are included, and each layer is related to the one below by a
$C'_{2z}$ rotation through the $C$ sites.
(c) Four-layer plaquette model. Same as (b), except that the rotation
is now $C'_{4z}$ instead.}
\label{fig:plaqmodel}
\end{figure}

The model studied here draws inspiration from a 2D square-plaquette model developed in Ref.~\cite{ceresoli-prb2006}. As presented there, the model consists of a nearest-neighbor TB Hamiltonian whose primitive cell consists of four sites labeled A-D as shown in Fig.~\ref{fig:plaqmodel}(a), with lattice vectors $\textbf{a}_1=\hat{\textbf{x}}$ and $\textbf{a}_2=\hat{\textbf{y}}$. Each of the four resulting plaquettes in the unit cell is threaded by a magnetic flux that corresponds to endowing some of the hopping amplitudes with an identical complex phase $e^{i\phi/2}$. The moduli of all the nonzero hoppings are set to an identical value $t$.

Fig.~\ref{fig:plaqmodel}(b) provides a detailed illustration of the model that we study. Our model features equidistantly spaced layers of the plaquette model stacked directly on top of one another along the $\hat{\textbf{z}}$ direction. Subsequent layers are related by a time-reversed rotation $C'_{2z}$ about a $z$ axis passing through the C site, resulting in two layers per unit cell with $\textbf{a}_3=\hat{\textbf{z}}$ forming the third lattice vector. The onsite energies are chosen to be $(E_A,E_B,E_C,E_D)=(\Delta,-\Delta,\Delta,-\Delta)$ for all layers, consistent with the rotational symmetry. Only interlayer A-A and C-C hoppings are included, with these alternating between $t_3$ and $t'_3$ for subsequent layers. Finally, the chosen flux pattern for the bottom layer is $(\Phi_1,\Phi_2,\Phi_3,\Phi_4)=(2\phi,-\phi,0,-\phi)$ and the rest of the nearest neighbors are coupled by real and nonzero hoppings $t$ (the modulus of the complex hoppings).  The parameter values are chosen such that the layers have a vanishing Chern number in the decoupled limit.

This model possesses a time-reversed inversion symmetry $I'$ about a point
located midway between C sites on neighboring layers,
as well as a two-fold rotation $C_{2x\overline{y}}$ about an axis
located between layers and passing
above A and C sites. Additionally, there are $m'_{xy}$ and
$m'_{x\overline{y}}$ mirror planes that pass through sites B and D, and A and
C, respectively. There is also a time reversed glide mirror plane $\{m'_z|\boldsymbol{\tau}/2\}$ passing in between layers with $\boldsymbol{\tau}=\hat{\textbf{x}}+\hat{\textbf{y}}$. These symmetries ensure that the bulk magnetization remains zero. 

The CSA coupling is tuned by keeping all parameters fixed except $t'_3$. If $t_3$ and $t'_3$
are equal, the model gains an axion-odd $\{C'_{2z}|c/2\}$ screw axis
passing through the C sites. 
The parameters for this model are chosen as $\Delta=-0.5$, $t=0.06$, $\phi=\pi/4$, $t_3=0.1$, and
$0.05\leq t'_3\leq0.15$.

\begin{table}[b]
\caption{\label{tab:2plaq_straight_mags_axeven}Surface magnetizations and hinge currents (in units of $10^{-6} e/\hbar$) for the two-layer plaquette model with straight-edge terminations and with $t'_3=0.15$ (axion-even).}
\begin{ruledtabular}
\begin{tabular}{lcccccc}
& $M_\perp^\text{B}$ & $M_\perp^\text{F}$ & $M_\perp^\text{L}$ &
$\Delta M^\text{FB}$ & $\Delta M^\text{BL}$ & $\Delta M^\text{LF}$ \\
\hline
$M^{\cal CC}$ & 2.3012 & 0.0319 & 0.0319 & $-$2.2693 & 2.2693 & 0.0000 \\ 
$M^{\cal EE}$ & 2.2995 & 0.0327 & 0.0327 & $-$2.2668 & 2.2668 & 0.0000 \\ 
$M^\text{lin}$ & 2.3001 & 0.0324 & 0.0324 & $-$2.2676 & 2.2676 & 0.0000 \\ 
$I_\text{calc}$ & & & & $-$2.2676 & 2.2676 & 0.0000
\end{tabular}
\end{ruledtabular}
\end{table}

We will consider two distinct surface terminations in the directions transverse to the layer stacking, and for each we will investigate the behavior of the resulting surface magnetizations and hinge currents. The first surface termination results from creating slabs along $\textbf{a}_1$ and $\textbf{a}_2$, and we refer to these as the straight-edge terminations. The second type of termination results from cutting along the $\textbf{a}_1\pm\textbf{a}_2$ directions, which we refer to as the zigzag termination. For our calculations with this termination, we construct supercells with lattice vectors $\textbf{A}_1=\textbf{a}_1-\textbf{a}_2$, $\textbf{A}_2=\textbf{a}_1+\textbf{a}_2$, and $\textbf{A}_3=\textbf{a}_3$, and cut slabs along the $\textbf{A}_1$ and $\textbf{A}_2$ directions.

\subsubsection{Two-layer plaquette model: straight edges}
\label{2plaq_straight}

In this case
$M^{\text{B}}_{\perp}$, $M^{\text{F}}_{\perp}$, and $M^{\text{L}}_{\perp}$ are found from slabs with eleven unit cells along $\textbf{a}_3$, seven unit cells along $\textbf{a}_2$, and seven unit cells along $\textbf{a}_1$, respectively. For each magnetization, a $7\times7$ $\textbf{k}$-space mesh in reduced coordinates is used.

The macroscopic hinge current $I^{\text{FB}}$ is found using an $x$-extensive rod composed of eleven unit cells along $\textbf{a}_3$ and seven unit cells along $\textbf{a}_2$. The $y$-extensive rod employed for $I^{\text{BL}}$ is composed of eleven unit cells along $\textbf{a}_3$ and seven unit cells along $\textbf{a}_1$, while the $z$-extensive rod for $I^{\text{LF}}$ is composed of seven unit cells along both $\textbf{a}_1$ and $\textbf{a}_2$.
For each rod, a 1D $\textbf{k}$-space sampling of thirty points is employed.

\begin{table}[b]
\caption{\label{tab:2plaq_zigzag_mags_axodd}Surface magnetizations and hinge currents (in units of $10^{-6} e/\hbar$) for the two-layer plaquette model with zigzag terminations and with $t'_3=0.1$ (axion-odd).}
\begin{ruledtabular}
\begin{tabular}{lcccccc}
& $M_\perp^\text{B}$ & $M_\perp^\text{F}$ & $M_\perp^\text{L}$ &
$\Delta M^\text{FB}$ & $\Delta M^\text{BL}$ & $\Delta M^\text{LF}$ \\
\hline
$M^{\cal CC}$ & 2.3768 & 0.0000 & 0.0000 & $-$2.3768 & 2.3768 & 0.0000 \\ 
$M^{\cal EE}$ & 2.3768 & 0.0000 & 0.0000 & $-$2.3768 & 2.3768 & 0.0000 \\ 
$M^\text{lin}$ & 2.3768 & 0.0000 & 0.0000 & $-$2.3768 & 2.3768 & 0.0000 \\ 
$I_\text{calc}$ & & & & $-$2.3768 & 2.3768 & 0.0000 
\end{tabular}
\end{ruledtabular}
\end{table}

For $t'_3=0.1$ and $t'_3=0.15$, respectively, Tables~\ref{tab:2plaq_straight_mags_axodd} and~\ref{tab:2plaq_straight_mags_axeven} report the values of the surface magnetizations computed from the different markers. From the tables, we see that for each individual marker, the computed values of $M^{\text{F}}_{\perp}$ and $M^{\text{L}}_{\perp}$ are identical, and therefore lead to a vanishing $\Delta M^{\text{LF}}$, which also agrees with the directly computed $I^{\text{LF}}$. This may be understood as a consequence of the $m'_{x\bar{y}}$ symmetry of the system, which eliminates the hinge current and maps $M^{\text{L}}_{\perp}$ and $M^{\text{F}}_{\perp}$ into each other.

For the particular case of the axion-odd regime of $t'_3=0.1$, $M^{\text{F}}_{\perp}$ and $M^{\text{L}}_{\perp}$ are not just identical for all markers, but zero as well, which can be understood as a consequence of the $m'_{x\bar{y}}$, $m'_{xy}$, and $\{C'_{2z}|c/2\}$ symmetries. The $m'_{x\bar{y}}$ mirror plane enforces identical $M_{\perp}$ values on surfaces with outward unit normals $-\hat{\textbf{x}}$ and $-\hat{\textbf{y}}$, as well as on the surfaces with outward unit normals $\hat{\textbf{x}}$ and $\hat{\textbf{y}}$. The $m'_{xy}$ symmetry subsequently implies that the $M_{\perp}$ values on all four surfaces are the same, while $\{C'_{2z}|c/2\}$ ensures that $M_{\perp}=0$.

In the axion-odd regime when $t'_3=0.1$, all markers yield the same value of magnetization at a single surface. However, in the axion-even regime when $t'_3=0.15$, the surface magnetizations derived from different markers do not agree, and furthermore only the $\mathcal{M}^{\text{lin}}$ marker correctly predicts the hinge currents. We will soon see that among the three markers, it appears that only $\mathcal{M}^{\text{lin}}$ consistently predicts the hinge currents, regardless of whether the system is in the axion-odd or axion-even regime.

\subsubsection{Two-layer plaquette model: zigzag edges}
\label{2plaq_zigzag}
In order to describe the slabs and rods that occur when zigzag terminations are present, we turn to using the supercell vectors $\textbf{A}_1$, $\textbf{A}_2$, and $\textbf{A}_3$ described at the beginning of the subsection. We align the $x$ and $y$ axes along $\textbf{A}_1$ and $\textbf{A}_2$ respectively. Since the bottom surface of a slab cut along $\textbf{A}_3$ is identical up to a rotation in the $x$-$y$ plane to the bottom surface of the slabs used to compute $M^{\text{B}}_{\perp}$ in the straight-edge case, the values of $M^{\text{B}}_{\perp}$ will be left unchanged. With the rotation of the coordinate axes, we also observe that the $m'_{x\bar{y}}$ and $m'_{xy}$ mirrors become $m'_x$ and $m'_y$ mirrors, respectively, the $C_{2x\bar{y}}$ axis becomes a $C_{2x}$ axis, and $\{m'_z|\boldsymbol{\tau}/2\}$ becomes $\{m'_z|\textbf{A}_2/2\}$.

In this case
$M^{\text{F}}_{\perp}$ and $M^{\text{L}}_{\perp}$ are found from slabs with four cells along $\textbf{A}_2$ and $\textbf{A}_1$, respectively. For each magnetization, a $7\times7$ $\textbf{k}$-space mesh in reduced coordinates is used. The macroscopic hinge current $I^{\text{FB}}$ is found from an $x$-extensive rod composed of nine cells along $\textbf{A}_3$ and nine cells along $\textbf{A}_2$. The $y$-extensive rod employed for $I^{\text{BL}}$ is composed of nine cells along $\textbf{A}_3$ and nine cells along $\textbf{A}_1$, while the $z$-extensive rod for $I^{\text{LF}}$ is composed of nine and nine cells along $\textbf{A}_1$ and $\textbf{A}_2$, respectively.
For each rod, a 1D $\textbf{k}$-space sampling of five points is employed.

\begin{table}[b]
\caption{\label{tab:2plaq_zigzag_mags_axeven}Surface magnetizations and hinge currents (in units of $10^{-6} e/\hbar$) for the two-layer plaquette model with zigzag terminations and with $t'_3=0.15$ (axion-even).}
\begin{ruledtabular}
\begin{tabular}{lcccccc}
& $M_\perp^\text{B}$ & $M_\perp^\text{F}$ & $M_\perp^\text{L}$ &
$\Delta M^\text{FB}$ & $\Delta M^\text{BL}$ & $\Delta M^\text{LF}$ \\
\hline
$M^{\cal CC}$ & 2.3012 & $-$0.0014 & 0.0650 & $-$2.3026 & 2.2362 & 0.0664 \\ 
$M^{\cal EE}$ & 2.2995 & $-$0.5697 & 0.6350 & $-$2.8692 & 1.6645 & 1.2047 \\ 
$M^\text{lin}$ & 2.3001 & $-$0.3802 & 0.4450 & $-$2.6803 & 1.8551 & 0.8252 \\ 
$I_\text{calc}$ & & & & $-$2.6803 & 1.8551 & 0.8252
\end{tabular}
\end{ruledtabular}
\end{table}

Tables \ref{tab:2plaq_zigzag_mags_axodd} and \ref{tab:2plaq_zigzag_mags_axeven} demonstrate the values of the surface magnetizations computed from the different markers for $t'_3=0.1$ and $t'_3=0.15$, respectively. In the axion-odd case, we see that for a fixed surface all the markers yield identical values of the magnetization. Furthermore, the values of $M^{\text{F}}_{\perp}$ and $M^{\text{L}}_{\perp}$ vanish due to the $m'_x$, $m'_y$, and $\{C'_{2z}|c/2\}$ symmetries. $m'_x$ ensures identical $M_{\perp}$ values on surfaces with outward unit normals $\hat{\textbf{x}}$ and $-\hat{\textbf{x}}$, while $m'_y$ does so for surfaces with outward unit normals $\hat{\textbf{y}}$ and $-\hat{\textbf{y}}$. $\{C'_{2z}|c/2\}$ then ensures that these surface magnetizations are zero.

\begin{figure*}[t]
\centering
\includegraphics[width=7.3in]{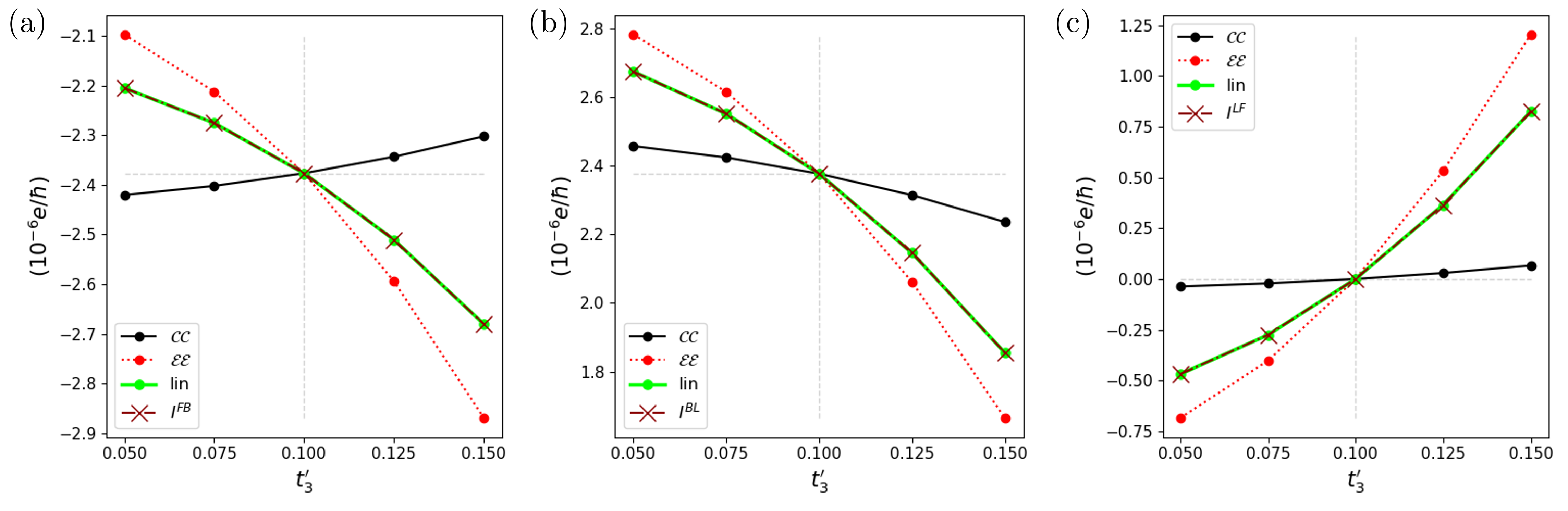}
\caption{Plots of the (a) $I^{\text{FB}}$, (b) $I^{\text{BL}}$, (c) $I^{\text{LF}}$ hinge currents for the two-layer square plaquette model with zigzag edges versus $t'_3$. The plots compare the hinge current predicted from the different local markers to the directly computed hinge current. When $t'_3=0.1$ (axion-odd regime), all the markers yield the correct values of the hinge currents, as indicated by the intersections of the dashed horizontal and vertical gray lines. Generally, only the surface magnetizations found from the `lin' marker always match the directly computed hinge currents.}
\label{fig:2plaq_zigzag_currents}
\end{figure*}

In the axion-even regime, the markers at a single surface facet yield differing values of the magnetizations, and none except the `lin' marker predict the hinge currents correctly. 

These observations are highlighted in Fig.~\ref{fig:2plaq_zigzag_currents}, which displays the hinge currents as a function of $t'_3$ and compares the currents predicted using the different markers to the directly computed hinge currents.

\subsection{Four-layer square plaquette model}
\label{4plaq}
The model studied in this section employs the same underlying square-plaquette layers that were used in Sec.~\ref{2plaq} and is depicted in Fig.~\ref{fig:plaqmodel}(c). Just like the model of Sec.~\ref{2plaq}, the present model features equidistantly spaced layers of the plaquette model stacked directly on top of one another along the $\hat{\textbf{z}}$ direction, but now the adjacent layers are related by a time-reversed rotation $C'_{4z}$ about a $z$ axis passing through the C site, resulting in four layers per unit cell. The lattice vector along the stacking direction is $\textbf{a}_3=\hat{\textbf{z}}$. Otherwise, the models share the same features.

\begin{table}[b]
\caption{\label{tab:4plaq_straight_mags_axodd}Surface magnetizations and hinge currents (in units of $10^{-7} e/\hbar$) for the four-layer plaquette model with straight-edge terminations and with $t'_3=0.1$ (axion-odd).}
\begin{ruledtabular}
\begin{tabular}{lcccccc}
& $M_\perp^\text{B}$ & $M_\perp^\text{F}$ & $M_\perp^\text{L}$ &
$\Delta M^\text{FB}$ & $\Delta M^\text{BL}$ & $\Delta M^\text{LF}$ \\
\hline
$M^{\cal CC}$ & 3.24982 & 0.00059 & $-$0.00059 & $-$3.24923 & 3.25040 & $-$0.00117 \\ 
$M^{\cal EE}$ & 3.24982 & 0.00066 & $-$0.00066 & $-$3.24915 & 3.25048 & $-$0.00133 \\ 
$M^\text{lin}$ & 3.24982 & 0.00064 & $-$0.00064 & $-$3.24918 & 3.25046 & $-$0.00128 \\ 
$I_\text{calc}$ & & & & $-$3.24918 & 3.25046 & $-$0.00128 
\end{tabular}
\end{ruledtabular}
\end{table}

This model
features a two-fold rotation axis $C_{2x}$ between the second and third
layers in the unit cell and passing above the C and D sites. There is
also a $\{C_{2z}|c/2\}$ screw axis that passes through the C sites.
Together these symmetries eliminate the bulk magnetization.
If the interlayer hoppings $t_3$ and $t'_3$ are equal, the model gains
an additional axion-odd $\{C'_{4z}|c/4\}$ screw axis running through the
C sites.
All parameters except $t'_3$ are kept fixed in order to access the
quantized CSA coupling regime. We employ the same values of the
parameters as in Sec.~\ref{2plaq}, but here we set the onsite energy $\Delta=-1$.

For this model, we also study the behavior of hinge currents and surface magnetizations for the different transverse surface terminations.

\subsubsection{Four-layer plaquette model: straight edges}
\label{4plaq_straight}
The magnetizations $M^{\text{B}}_{\perp}$, $M^{\text{F}}_{\perp}$, and $M^{\text{L}}_{\perp}$ are found from slabs with six unit cells along $\textbf{a}_3$, four unit cells along $\textbf{a}_2$, and four unit cells along $\textbf{a}_1$, respectively. For each magnetization, a $7\times7$ $\textbf{k}$-space mesh in reduced coordinates is used.

\begin{table}[b]
\caption{\label{tab:4plaq_straight_mags_axeven}Surface magnetizations and hinge currents (in units of $10^{-7} e/\hbar$) for the four-layer plaquette model with straight-edge terminations and with $t'_3=0.15$ (axion-even).}
\begin{ruledtabular}
\begin{tabular}{lcccccc}
& $M_\perp^\text{B}$ & $M_\perp^\text{F}$ & $M_\perp^\text{L}$ &
$\Delta M^\text{FB}$ & $\Delta M^\text{BL}$ & $\Delta M^\text{LF}$ \\
\hline
$M^{\cal CC}$ & 3.22067 & 0.01107 & 0.00912 & $-$3.20960 & 3.21155 & $-$0.00195 \\ 
$M^{\cal EE}$ & 3.22047 & 0.01130 & 0.00910 & $-$3.20916 & 3.21137 & $-$0.00221 \\ 
$M^\text{lin}$ & 3.22053 & 0.01123 & 0.00911 & $-$3.20931 & 3.21143 & $-$0.00212 \\ 
$I_\text{calc}$ & & & & $-$3.20931 & 3.21143 & $-$0.00212
\end{tabular}
\end{ruledtabular}
\end{table}

The macroscopic hinge current $I^{\text{FB}}$ is found using an $x$-extensive rod composed of six unit cells along $\textbf{a}_3$ and four unit cells along $\textbf{a}_2$. The $y$-extensive rod employed for $I^{\text{BL}}$ is composed of six unit cells along $\textbf{a}_3$ and four unit cells along $\textbf{a}_1$, while the $z$-extensive rod for $I^{\text{LF}}$ is composed of four unit cells along both $\textbf{a}_1$ and $\textbf{a}_2$.
For each rod, a 1D $\textbf{k}$-space sampling of thirty points is employed.

For $t'_3=0.1$ and $t'_3=0.15$, respectively, Tables~\ref{tab:4plaq_straight_mags_axodd} and~\ref{tab:4plaq_straight_mags_axeven} report the values of the surface magnetizations computed from the different markers. In the axion-odd regime when $t'_3=0.1$, all markers yield the same value for one of the surface magnetizations, namely $M_\perp^\text{B}$.  This time, however, for both $M_\perp^\text{F}$ and $M_\perp^\text{L}$, the markers yield different values. Moreover, only the `lin' marker accurately predicts the hinge currents.  For each marker, the values for $M_\perp^\text{F}$ and $M_\perp^\text{L}$ are equal and opposite, which may be understood to be a consequence of the $\{C'_{4z}|c/4\}$ screw symmetry, see Fig.~\ref{fig:screws}(b). Nevertheless, we now have a case where the markers disagree, and where only one of them correctly predicts the hinge currents.

\begin{table}[b]
\caption{\label{tab:4plaq_zigzag_mags_axodd}Surface magnetizations and hinge currents (in units of $10^{-7} e/\hbar$) for the four-layer plaquette model with zigzag terminations and with $t'_3=0.1$ (axion-odd).}
\begin{ruledtabular}
\begin{tabular}{lcccccc}
& $M_\perp^\text{B}$ & $M_\perp^\text{F}$ & $M_\perp^\text{L}$ &
$\Delta M^\text{FB}$ & $\Delta M^\text{BL}$ & $\Delta M^\text{LF}$ \\
\hline
$M^{\cal CC}$ & 3.24982 & 0.00000 & 0.00000 & $-$3.24982 & 3.24982 & 0.00000 \\ 
$M^{\cal EE}$ & 3.24982 & 0.00000 & 0.00000 & $-$3.24982 & 3.24982 & 0.00000 \\ 
$M^\text{lin}$ & 3.24982 & 0.00000 & 0.00000 & $-$3.24982 & 3.24982 & 0.00000 \\ 
$I_\text{calc}$ & & & & $-$3.24982 & 3.24982 & 0.00000
\end{tabular}
\end{ruledtabular}
\end{table}

\begin{figure}[t]
\centering
\includegraphics[width=2.4in]{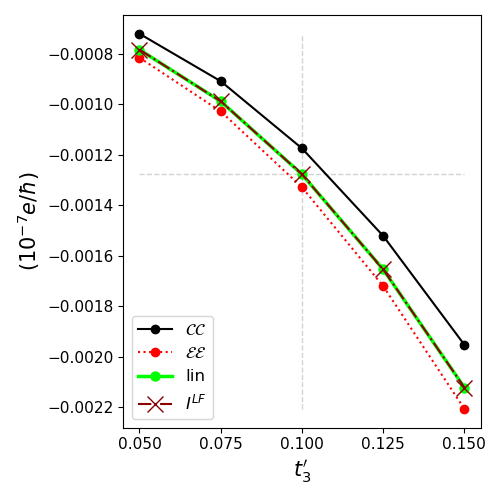}
\caption{Plot of the hinge current $I^{\text{LF}}$ as a function of $t'_3$ for the four-layer square plaquette model with straight edges. In the axion-odd regime when $t'_3=0.1$, only the `lin' marker correctly predicts the hinge current; this is also the case even when no axion-odd symmetry is present.}
\label{fig:Iz_4plaq_straight}
\end{figure}

The same is true in the axion-even regime of Table~\ref{tab:4plaq_straight_mags_axeven}.
Fig.~\ref{fig:Iz_4plaq_straight} demonstrates these facts for the prediction of the $I^{\text{LF}}$ hinge current. For all the values of $t'_3$ that we used, covering both axion-odd and axion-even regimes, only the `lin' marker was consistent in its prediction of the current.

\subsubsection{Four-layer plaquette model: zigzag edges}
\label{4plaq_zigzag}

As in the case of two-layer plaquette model with zigzag terminations in Sec.~\ref{2plaq_zigzag}, we turn to using the supercell vectors $\textbf{A}_1=\textbf{a}_1-\textbf{a}_2$, $\textbf{A}_2=\textbf{a}_1+\textbf{a}_2$, and $\textbf{A}_3=\textbf{a}_3$ for our calculations involving the zigzag edges of the four-layer plaquette model. We align the $x$ and $y$ axes along $\textbf{A}_1$ and $\textbf{A}_2$ respectively. Since the bottom surface of a slab cut along $\textbf{A}_3$ is identical up to a rotation in the $x$-$y$ plane to the bottom surface of the slabs used to compute $M^{\text{B}}_{\perp}$ in the straight-edge case, the values of $M^{\text{B}}_{\perp}$ will be left unchanged. With the rotation of the coordinate axes, $C_{2x}$ becomes $C_{2xy}$. $M^{\text{F}}_{\perp}$ and $M^{\text{L}}_{\perp}$ in this case are found from slabs with three cells along $\textbf{A}_2$ and $\textbf{A}_1$, respectively. For each magnetization, a $5\times5$ $\textbf{k}$-space mesh is used in reduced coordinates.

The macroscopic hinge current $I^{\text{FB}}$ is found from an $x$-extensive rod composed of four unit cells along $\textbf{A}_3$ and six unit cells along $\textbf{A}_2$. The $y$-extensive rod employed for $I^{\text{BL}}$ is composed of four unit cells along $\textbf{A}_3$ and six unit cells along $\textbf{A}_1$, while the $z$-extensive rod for $I^{\text{LF}}$ is composed of six cells along both $\textbf{A}_1$ and $\textbf{A}_2$.
For each rod, a 1D $\textbf{k}$-space sampling of 5 points is employed.

Tables~\ref{tab:4plaq_zigzag_mags_axodd} and~\ref{tab:4plaq_zigzag_mags_axeven} report the values of the surface magnetizations computed from the different markers for $t'_3=0.1$ and $t'_3=0.15$, respectively.
This time,
in the axion-odd regime when $t'_3=0.1$, we see that for any given surface facet, all markers yield the same value of magnetization, and when compared across facets, all accurately predict the hinge currents. The values of $M_\perp^\text{F}$ and $M_\perp^\text{L}$ vanish, which may be understood to be a consequence of the simultaneous $C_{2xy}$ and $\{C'_{4z}|c/4\}$ symmetries. The former symmetry implies that $M_\perp^\text{F}=M_\perp^\text{L}$, while the latter results in both vanishing. 

\begin{table}[b]
\caption{\label{tab:4plaq_zigzag_mags_axeven}Surface magnetizations and hinge currents (in units of $10^{-7} e/\hbar$) for the four-layer plaquette model with zigzag terminations and with $t'_3=0.15$ (axion-even).}
\begin{ruledtabular}
\begin{tabular}{lcccccc}
& $M_\perp^\text{B}$ & $M_\perp^\text{F}$ & $M_\perp^\text{L}$ &
$\Delta M^\text{FB}$ & $\Delta M^\text{BL}$ & $\Delta M^\text{LF}$ \\
\hline
$M^{\cal CC}$ & 3.22067 & 0.01023 & 0.01023 & $-$3.21045 & 3.21045 & 0.00000 \\ 
$M^{\cal EE}$ & 3.22047 & 0.01033 & 0.01033 & $-$3.21014 & 3.21014 & 0.00000 \\ 
$M^\text{lin}$ & 3.22053 & 0.01029 & 0.01029 & $-$3.21024 & 3.21024 & 0.00000 \\ 
$I_\text{calc}$ & & & & $-$3.21024 & 3.21024 & 0.00000
\end{tabular}
\end{ruledtabular}
\end{table}

In the axion-even regime when $t'_3=0.15$, we find a result that is more reminiscent of the straight-edge case. That is, even for a single surface facet, the markers yield different values of the magnetization, and the hinge currents computed from the differences of $M_{\perp}$ values disagree with each other when computed using different markers.  They also disagree with the direct calculation of the hinge current except for one case, namely that of the $\mathcal{M}^{\text{lin}}$ marker. $M^{\text{F}}_{\perp}$ and $M^{\text{L}}_{\perp}$ are found to be identical, as the $C_{2xy}$ symmetry is still present in this setup. However, only the $\mathcal{M}^{\text{lin}}$ marker yields magnetizations that correctly predict the hinge currents.

\subsubsection{Summary}

For the four-layer plaquette model, unlike the models studied previously,
we find that different markers generally disagree on the values of $M_\perp$ and on their
predictions for the hinge currents.  This occurs in both the axion-odd and axion-even
cases for straight-edge surface terminations, but only in the axion-even case for
zigzag surface terminations.  Only the $\mathcal{M}^{\text{lin}}$ marker consistently predicts the correct hinge currents.

\begin{table}[b]
\caption{\label{tab:CubeResultsDiag}Diagonal elements of the traceless marker-based MQMs compared to the MQM from the current-based formula. Only the MQM derived from the $\mathcal{M}^{\text{lin}}$ marker matches the current based quadrupole. Entries are in units of $e/\hbar$.}
\begin{ruledtabular}
\begin{tabular}{lccc}
& $Q_{xx}$ & $Q_{yy}$ & $Q_{zz}$  \\
\hline
$\cal CC$ & $-9.9765\times10^{-3}$ & $-9.9765\times10^{-3}$ & $ 1.9953\times10^{-2}$ \\ 
$\cal EE$ & $-1.2540\times10^{-2}$ & $-1.2540\times10^{-2}$ & $2.5079\times10^{-2}$  \\ 
$\text{lin}$ & $-1.1685\times10^{-2}$ & $-1.1685\times10^{-2}$ & $2.3370\times10^{-2}$  \\ 
$\text{Curr.}$ & $-1.1685\times10^{-2}$ & $-1.1685\times10^{-2}$ & $2.3370\times10^{-2}$ 
\end{tabular}
\end{ruledtabular}
\end{table}

\subsection{Local markers and magnetic quadrupole moment of a finite system}
\label{MolecQuadMom}

\begin{figure}[t]
\centering
\includegraphics[width=2.7in]{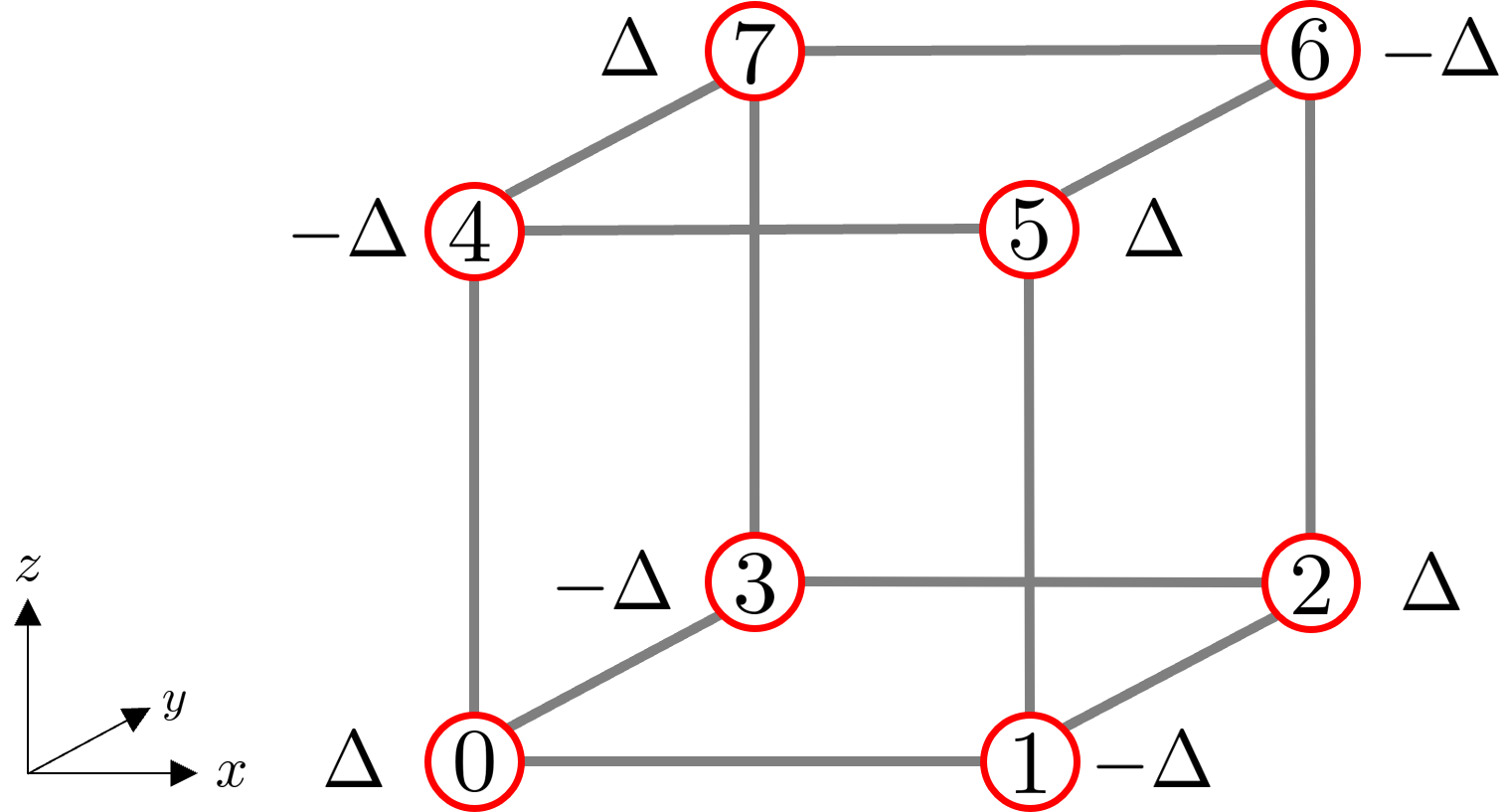}
\caption{Illustration of the structure, labeling, and onsite energies of the cubic molecule model discussed in Sec.~\ref{MolecQuadMom}.}
\label{fig:cubemodel}
\end{figure}

We now return to investigate the questions posed in Sec.~\ref{MQM_marker} by conducting a numerical study of a spinless TB model of a cubic \say{molecule}.
The system is designed to have a symmetry that enforces a vanishing magnetic dipole moment, but otherwise is as arbitrary as possible. In particular, it has no axion-odd symmetries. The structure, TB site labeling, and onsite energies of the model are depicted in Fig.~\ref{fig:cubemodel}. Additionally, the model features hoppings $t_{ij}$  to nearest, second-nearest, and third-nearest neighbors, where the subscripts indicate a hopping from TB site $i$ to TB site $j$. The hoppings are generally complex with $t_{ij}=t^*_{ji}$ to ensure that the Hamiltonian is Hermitian.

\begin{table}[b]
\caption{\label{tab:CubeResultsOffDiag}Off diagonal elements of the traceless marker-based MQMs compared to the MQM from the current-based formula. Only the MQM derived from the $\mathcal{M}^{\text{lin}}$ marker matches the current-based quadrupole. Entries are in units of $e/\hbar$.}
\begin{ruledtabular}
\begin{tabular}{lcc}
& $Q_{xy}$ & $Q_{yx}$   \\
\hline
$\cal CC$ & $-6.4152\times10^{-2}$ & $6.4152\times10^{-2}$  \\ 
$\cal EE$ & $-8.4555\times10^{-2}$ & $8.4555\times10^{-2}$  \\ 
$\text{lin}$ & $-7.7754\times10^{-2}$ & $7.7754\times10^{-2}$  \\ 
$\text{Curr.}$ & $-7.7754\times10^{-2}$ & $7.7754\times10^{-2}$ 
\end{tabular}
\end{ruledtabular}
\end{table}

The Hamiltonian is chosen to obey an $S'_{4z}=m_zC'_{4z}$ improper rotation symmetry, where the $m_z$ mirror plane bisects the molecule, while the $C'_{4z}$ axis runs through the center of the cube. The resulting magnetic point group of the system is then $\{E,S'_{4z},(S'_{4z})^{-1},C_{2z}\}$, where $E$ is the identity operation and $C_{2z}$ is the square of $S'_{4z}$. These symmetries ensure a vanishing magnetic dipole moment, resulting in an origin-independent, physically meaningful MQM for the system. The magnetic point group also ensures that the MQM features $Q_{xx}=Q_{yy}$ and $Q_{xy}=-Q_{yx}$. Additionally, $Q_{zz}=-2Q_{xx}$, as the tensor is traceless.

In addition to the onsite energies, the Hamiltonian is parametrized by eight hopping parameters. The nearest neighbor hopping terms $t_1$, $t_2$, and $t_3$ parametrize, respectively, the hoppings
\begin{eqnarray*}
    t_{10}=t_{56}=t_{32}=t_{74},\\t_{30}=t_{54}=t_{12}=t_{76},\\t_{40}=t_{51}=t_{62}=t_{73}.
\end{eqnarray*} The second nearest neighbor hopping terms $t_4$, $t_5$, $t_6$, and $t_7$ parametrize the hoppings \begin{eqnarray*}
    t_{20}&=&t_{57},\\t_{31}&=&t_{64},\\t_{50}&=&t_{52}=t_{72}=t_{70},\\t_{14}&=&t_{16}=t_{36}=t_{34},
\end{eqnarray*} respectively. Finally, the third nearest neighbor hopping $t_8$ parametrizes the hoppings $t_{35}=t_{17}=t_{24}=t_{06}$. 

Generally, the hopping parameters are set to be complex, but due to the presence of the $C_{2z}$ symmetry in the magnetic point group, $t_4$ and $t_5$ must be real. The parameters for this model are chosen as $\Delta=-1$, $t_1=1+0.2i$, $t_2=1+0.3i$, $t_3=1-0.1i$, $t_4=-0.3$, $t_5=0.5$, $t_6=-0.2+0.15i$, $t_7=0.3-0.25i$, and $t_8=0.4+0.6i$.

We then calculate the current-based quadrupole and the traceless dipole-based quadrupoles based on the $\mathcal{CC}$, $\mathcal{EE}$ and `lin' markers. Table~\ref{tab:CubeResultsDiag} lists the results for diagonal elements of the quadrupoles, while Table~\ref{tab:CubeResultsOffDiag} lists the nonzero off diagonal elements. We see that all feature $Q_{xx}=Q_{yy}=-Q_{zz}/2$ and $Q_{xy}=-Q_{yx}$ as expected. Significantly, though, only the $\mathcal{M}^{\text{lin}}$ marker generates a quadrupole that matches the current-based MQM. 

\section{Discussion}
\label{Discussion}

The results presented above help us to address the main questions
motivating this work.  First, does the introduction of quantum
mechanics in the form of a marker-based theory provide a
prescription for computing $M_\perp$ directly for a given facet,
in such a way that hinge currents at adjoining facets are correctly
predicted?  Is this equally true in the axion-odd and axion-even cases?
Second, if multiple markers succeed in doing so, does the quantum
theory inherit the shift freedom of the classical theory?
The most simplistic scenario would be one in which all of the physically
acceptable markers identified at the end of Sec.~\ref{Markers}
yield identical $M_\perp$ values for any given facet over all models,
and correctly predict the hinge currents where facets meet.

Focusing first on the axion-odd case, we found that our tests were
consistent with this simplistic scenario for the alternating Haldane
and two-layer square-plaquette models.  However, the four-layer
square-plaquette model does show signs of trouble.  For the
straight-edge surface terminations,
the magnetizations found from
the $\mathcal{M}^{\mathcal{C},\mathcal{C}}$,
$\mathcal{M}^{\mathcal{E},\mathcal{E}}$, and
$\mathcal{M}^{\text{lin}}$ markers do not agree with each other.
Furthermore, among these, we found that only the $\mathcal{M}^{\text{lin}}$
marker correctly predicts the hinge current.  Thus, while the
physical magnetization is uniquely defined in the axion-odd case,
only the $\mathcal{M}^{\text{lin}}$ marker consistently
gives the correct value for it in our tests.

When tuning from the axion-odd to the axion-even regime,
the same conclusion was reinforced.
While all markers continued to agree with each other
for the stacked Haldane model, they disagreed in all other cases,
i.e., for the two-layer square-plaquette model and all
surface terminations of the four-layer square-plaquette model.
Again, the only marker that always reproduced the hinge currents
across all models was the $\mathcal{M}^{\text{lin}}$ marker.

The importance
of the $\mathcal{M}^{\text{lin}}$ marker was also highlighted by
our numerical finding that the dipole-based traceless MQM derived
from it appears to correctly reproduce the current-based MQM in
finite systems.  Thus, we provisionally identify $M^\mathrm{lin}$
as the best candidate for a marker-based formulation of a theory
of surface orbital magnetization.

Given the $\mathcal{M}^{\text{lin}}$ marker's evident connection
to the MQM, we have also explored the symmetries of our models
to understand whether they
permit a nonzero bulk MQM.  While the question of defining an
orbital MQM for a general bulk material has received some attention
\cite{shitade-prb2018,gao-prb2018}, the situation appears to
remain unsettled. Assuming such a definition is possible, however,
it must respect the symmetries of the magnetic point group.

The four-layer square-plaquette model is most informative in
this respect.
Focusing first on the axion-odd case, the tensor is
diagonal with $Q_{xx}=-Q_{yy}$ nonzero and $Q_{zz}=0$ when
the coordinate axes are chosen as in Fig.~\ref{fig:plaqmodel}(c)
for the straight-edge surfaces.  In the zigzag-edge case,
we rotated the axes about $\hat{\textbf{z}}$ by $-\pi/4$,
transforming the tensor such that the only nonzero elements
were $Q'_{xy}=Q'_{yx}$.  We found that the markers disagreed with
each other in precisely those cases in which the corresponding
diagonal MQM tensor element was nonzero, namely for the side surfaces in the straight-edge
case.  According to Eq.~(\ref{eq:MQM_surfMag}), these were also
the cases in which the bulk MQM has the right symmetry
to contribute to the surface magnetization.  When the axion-odd
symmetry was broken, we found that all three diagonal elements
were present in the MQM for both surface terminations, and that
the markers disagreed with each other for all surfaces

For the alternating Haldane and two-layer square-plaquette
models, the symmetry of the MQM is such that it vanishes in the
axion-odd regime but acquires some diagonal elements in the
axion-even regime. 
For the latter model, the markers also failed to agree in
the axion-even regime.

To summarize our discussion of the numerical results, when inspecting all three models, we
found that the only cases in which markers disagree, with only the
$\mathcal{M}^{\text{lin}}$ marker correctly predicting the hinge currents,
were those in which diagonal elements of the MQM tensor were present.
Thus, however the bulk MQM might be defined, our
results suggest that only the $\mathcal{M}^{\text{lin}}$ marker
accurately takes into account its contribution to surface
magnetization.

Notably, our findings
provide no evidence for the possible marker shift freedom
of the surface magnetizations within the quantum-mechanical marker-based theory. In fact, since only the $\mathcal{M}^{\text{lin}}$ marker always yielded the correct hinge currents, we do not have multiple marker candidates as would be be needed to allow us to test this proposition.

We find ourselves, then, in a somewhat ambiguous position.  Our
empirical evidence, based on the model studies, indicates that the original
Bianco-Resta marker $\mathcal{M}^\mathcal{C,C}$ is not suitable for
coarse-graining to obtain the surface magnetization.  Instead, a modification
of it, namely $\mathcal{M}^\mathrm{lin}$, has passed all tests and appears to be a suitable marker. Still, despite a concerted effort to find a formal connection, we do not yet have a fundamental understanding as to why $M^\mathrm{lin}$ succeeds where the others do not. It also cannot yet be ruled out that $M^\mathrm{lin}$ does not pass all tests in models beyond those considered in this paper. Our work thus raises these issue as being of paramount importance to be addressed by future theoretical investigations.

We now also comment in more detail on the connection of this work to that of Zhu et al.~\cite{zhu-prb2021}. The authors of that work also introduced the notion of surface orbital magnetization, and similarly used the local marker formulation of orbital magnetism to compute it. Their primary focus, as we mentioned just before Sec.~\ref{orig}, was to explore the ``facet magnetic compressibility" $dM_\perp/d\mu$ that is directly proportional to the geometric component of the surface AHC, especially in the context of links to higher-order topology. We agree with their conclusion that this compressibility is uniquely determined. On the other hand, where they assumed that the Bianco-Resta $\mathcal{M}^\mathcal{C,C}$ marker could be applied to define surface magnetization, our findings indicate that this assumption was problematic.  We also note that they
did not explore the issue of a possible shift freedom of the surface magnetization.

Of course, it would be desirable to make contact with experiment.
We have argued that the hinge currents are observable in principle via the
magnetic fields they generate, which could be observed by scanning magnetic probes.
The surface magnetization $M_\perp$ itself in the interior of a facet does not produce any observable electric or magnetic potential, and so cannot be observed directly. It is possible that optical probes, such as terahertz Kerr reflectivity, might provide information.  However, one should keep in mind that when symmetry allows for the presence of a surface orbital magnetization, it also allows for a surface spin magnetization, which is likely to be much larger. Thus, filtering out just the orbital component may be challenging.

Other generalizations of our present work remain to be developed. The
symmetry analysis of Sec.~\ref{Symms} was conducted in full generality
and is applicable to insulating as well as metallic systems at any temperature. However, it is not immediately clear whether it is possible to compute surface magnetization using the local marker for $T\neq0$.
A crucial assumption going into the construction of the local marker is the idempotency of $P$ \cite{bianco-prl2013}; in the case of finite temperature, $P$ no longer displays this behavior. Additionally, for metals at $T=0$,
even though $P$ remains idempotent, it is no longer exponentially localized, with $P(\textbf{r},\textbf{r}')$ featuring power law decay with $|\textbf{r}-\textbf{r}'|$. It is not immediately clear how this affects the local marker and its subsequent coarse-grained average. In these situations, it appears likely that a different method of computing surface magnetization will be needed. We note, however, that prior numerical tests performed on TB models of metals at $T=0$ suggest that the local marker is able to reproduce the bulk magnetization of Eq.~(\ref{eq:MarkerEq}) within open boundary conditions, but exhibits slower convergence with sample size than in the insulating case \cite{marrazzo-prl2016}.

Additional directions
to explore include the presence of facet magnetizations for systems with
bulk magnetization, as well as for axion-odd systems with
$\theta_{\text{CS}}=\pi$. It would also be interesting to see whether a Wannier-based formulation for surface magnetization is possible, at least in
insulators. This is particularly attractive as the theories of bulk electric polarization, bulk orbital magnetization, and edge electric polarization can all be developed using a Wannier-based approach \cite{king-smith-prb1993,vanderbilt-prb1993,thonhauser-prl2005,ceresoli-prb2006,ren-prb2021,trifunovic-prresearch2020}.
Finally, we mention the prospects for a first-principles density-functional theory (DFT) implementation of the calculation of surface magnetization for topologically trivial bulk materials. The output of any DFT calculation may be adapted to a TB framework via Wannier interpolation, which may be accomplished by code packages such as \verb|Wannier90| \cite{Pizzi2020}. It is important to keep in mind, however, that in our calculations of the local markers we employed the diagonal approximation for the position operator. In the case of a Wannier interpolation, the position operator is not necessarily diagonal in the basis of Wannier states, and the off-diagonal terms must be kept in mind when performing calculations.

\section{Summary}
\label{Summary}

In this work, we have explored the possibility of defining surface orbital magnetization for insulating systems that are topologically trivial in the bulk and feature no bulk orbital magnetization. We have demonstrated that in a general classical context, the knowledge of the macroscopic currents residing on a hinge formed by two surface facets is sufficient to determine only the difference of the magnetizations of the two facets. The said macroscopic hinge currents are the physical observables corresponding to the presence of surface orbital magnetization, and this fact indicates that differences of surface orbital magnetization are observables as well. By means of a symmetry analysis, we have shown that individual values of facet magnetizations should be well defined when the bulk symmetry group contains an axion-odd symmetry. To develop the theory of surface magnetization in a quantum context, we further expanded on the local marker formulation of orbital magnetization developed by Bianco and Resta, and introduced a set of markers satisfying a list of physically meaningful transformation requirements. The coarse-grained averages of these markers were then used to compute surface magnetizations.

We tested the conclusions of our symmetry analysis and our formalism for computing surface magnetization on a series of spinless TB models.
We found that the markers do not always agree on the value of the surface magnetization,
even in the axion-odd case where we know from symmetry arguments that
$M_\perp$ is well defined.  Our results indicate that only a single marker on our
list, the $\mathcal{M}^{\text{lin}}$ marker, consistently predicts the correct
hinge currents.  According to the symmetry considerations
of Sec.~\ref{Symms}, in the axion-odd case we can then conclude that it
has correctly computed the unique $M_\perp$ values.  In the axion-even case, instead,
we can no longer say that it predicts uniquely correct $M_\perp$
values, only correct differences of $M_\perp$ values.

We additionally tested the markers to understand whether any of them correctly generated the current-based MQM of Eq.~(\ref{eq:MQM}), which we argued may contribute to surface magnetization. We found that only the $\mathcal{M}^{\text{lin}}$ marker correctly reproduced the current-based MQM when applied to a finite system with a nontrivial MQM and vanishing magnetic dipole moment. We also observed that
when the different markers failed to agree on the surface magnetizations for axion-odd systems, the bulk symmetry group permitted an MQM that could contribute to the surface magnetizations.

Overall, our study indicates that a coarse-graining of the $\mathcal{M}^{\text{lin}}$ marker provides a suitable framework for computing surface magnetizations. However, a formal derivation that would explain the unique success of the $\mathcal{M}^{\text{lin}}$ marker remains elusive, and questions persist about the definition and role of the bulk quadrupole tensor. Thus, while our work establishes a foundation for a theory of surface orbital magnetization, it also highlights the need for further work to resolve some important remaining questions.

\textbf{Note added:} After submission of this paper, we became aware of
a subsequent preprint by Gliozzi, Lin and Hughes \cite{gliozzi-arxiv2022} that also considers, though from a different perspective, issues related to surface orbital magnetization, hinge currents, and bulk magnetic quadrupole moments.

\acknowledgments
This work was supported by NSF Grants DMR-1954856 and DGE-1842213.

\appendix

\section{Local marker for slab geometries in the tight-binding representation}
\label{AppA}
In this Appendix, we provide further exposition on the calculation of the local markers in the context of TB models. For concreteness, we will focus on computing the marker for insulating slabs that are of sufficient finite thickness in the $z$ direction, but infinite in-plane, as in Sec.~\ref{AvgMark}.

The $\mathcal{M}^{\mathcal{C},\mathcal{C}}$, $\mathcal{M}^{\mathcal{E},\mathcal{E}}$, and $\mathcal{M}^{\text{lin}}$ local markers for the magnetization are all expressible in terms of the LC local markers of Eqs.~(\ref{eq:LCCo}-\ref{eq:LCR}) and their IC marker analogues. In this Appendix, we restrict ourselves to a calculation of the marker of Eq.~(\ref{eq:LCCo}), which for a TB site $i$ is given by \begin{equation} \mathcal{M}^{\mathcal{C}}_{\text{LC}}(i)=-\frac{e}{\hbar}\text{Im}\langle i|XHY^{\dagger}|i\rangle.
\end{equation} The calculations for all other LC and IC markers are similar, and we will simply state the results at the end of this Appendix. 

We remind the reader that in all of our models, we have set the Fermi energy inside the band gap at zero, so that we ignore the Chern marker contribution to the local marker. For details on how to compute the Chern marker in the TB representation, we refer the reader to Ref.~\cite{varnava-prb2018}.

We denote the valence ($v$) and conduction ($c$) band eigenstates of the slab Hamiltonian as $\psi_{v\textbf{k}}(\textbf{r})$ and $\psi_{c\textbf{k}}(\textbf{r})$, respectively, where $\textbf{k}=(k_x,k_y)$. The valence and conduction band projectors are then written as $P=(1/N_{\textbf{k}})\sum_{v\textbf{k}}|\psi_{v\textbf{k}}\rangle\langle\psi_{v\textbf{k}}|$ and $Q=(1/N_{\textbf{k}})\sum_{c\textbf{k}}|\psi_{c\textbf{k}}\rangle\langle\psi_{c\textbf{k}}|$, where $N_{\textbf{k}}$ is the number of $\textbf{k}$ points in the 2D Brillouin zone mesh. With this in mind, we observe that
\begin{widetext}
\begin{align}
\mathcal{M}^{\mathcal{C}}_{\text{LC}}(i)&=-\frac{e}{\hbar}\text{Im}\langle i|XHY^{\dagger}|i\rangle\nonumber\\&=
-\frac{e}{N_{\textbf{k}}\hbar}\text{Im}\sum_{\textbf{k}}\sum_{vv'cc'}\langle i|\psi_{v\textbf{k}}\rangle \langle\psi_{v\textbf{k}}|x|\psi_{c\textbf{k}}\rangle\langle\psi_{c\textbf{k}}|H|\psi_{c'\textbf{k}}\rangle \langle\psi_{c'\textbf{k}}|y|\psi_{v'\textbf{k}}\rangle\langle\psi_{v'\textbf{k}}|i\rangle\nonumber\\
&=-\frac{e}{N_{\textbf{k}}\hbar}\text{Im}\sum_{\textbf{k}}\sum_{vv'c}E_{c\textbf{k}}\psi_{v\textbf{k}}(i)\psi^*_{v'\textbf{k}}(i)X_{vc\textbf{k}}Y^{\dagger}_{cv'\textbf{k}},
\end{align}
\end{widetext}
where $E_{n\textbf{k}}$ is the eigenenergy corresponding to $|\psi_{n\textbf{k}}\rangle$, $X_{vc\textbf{k}}=\langle\psi_{v\textbf{k}}|x|\psi_{c\textbf{k}}\rangle$, and $Y^{\dagger}_{cv\textbf{k}}=\langle\psi_{c\textbf{k}}|y|\psi_{v\textbf{k}}\rangle$. 

In our calculations of the local markers, we do not directly compute the matrix elements of the position operators $x$ and $y$ in the energy eigenbasis; rather, we use the velocity operator $\hat{\textbf{v}}$, defined as \begin{equation}
    \hat{\textbf{v}}=\frac{1}{i\hbar}[\hat{\textbf{r}},H],
\end{equation} to rewrite \begin{equation}X_{vc\textbf{k}}=\frac{\langle\psi_{v\textbf{k}}|i\hbar v_x|\psi_{c\textbf{k}}\rangle}{E_{c\textbf{k}}-E_{v\textbf{k}}}\label{eq:X}\end{equation} and similarly for $Y^{\dagger}_{cv'\textbf{k}}$. 

The three LC markers are then \begin{gather}\mathcal{M}^{\mathcal{C}}_{\text{LC}}(i)=-\frac{e}{N_{\textbf{k}}\hbar}\text{Im}\sum_{\textbf{k}}\sum_{vv'c}E_{c\textbf{k}}\psi_{v\textbf{k}}(i)\psi^*_{v'\textbf{k}}(i)X_{vc\textbf{k}}Y^{\dagger}_{cv'\textbf{k}},\label{eq:LC_C}\\
\mathcal{M}^{\mathcal{L}}_{\text{LC}}(i)=-\frac{e}{N_{\textbf{k}}\hbar}\text{Im}\sum_{\textbf{k}}\sum_{cc'v}E_{c\textbf{k}}\psi_{c\textbf{k}}(i)\psi^*_{c'\textbf{k}}(i)X_{vc'\textbf{k}}Y^{\dagger}_{cv\textbf{k}},\label{eq:LC_L}\\
\mathcal{M}^{\mathcal{R}}_{\text{LC}}(i)=-\frac{e}{N_{\textbf{k}}\hbar}\text{Im}\sum_{\textbf{k}}\sum_{cc'v}E_{c'\textbf{k}}\psi_{c\textbf{k}}(i)\psi^*_{c'\textbf{k}}(i)X_{vc'\textbf{k}}Y^{\dagger}_{cv\textbf{k}}.\label{eq:LC_R}
\end{gather}
The three IC markers are \begin{gather}\mathcal{M}^{\mathcal{C}}_{\text{IC}}(i)=\frac{e}{N_{\textbf{k}}\hbar}\text{Im}\sum_{\textbf{k}}\sum_{cc'v}E_{v\textbf{k}}\psi_{c\textbf{k}}(i)\psi^*_{c'\textbf{k}}(i)X^{\dagger}_{cv\textbf{k}}Y_{vc'\textbf{k}},\label{eq:IC_C}\\
\mathcal{M}^{\mathcal{L}}_{\text{IC}}(i)=\frac{e}{N_{\textbf{k}}\hbar}\text{Im}\sum_{\textbf{k}}\sum_{vv'c}E_{v\textbf{k}}\psi_{v\textbf{k}}(i)\psi^*_{v'\textbf{k}}(i)X^{\dagger}_{cv'\textbf{k}}Y_{vc\textbf{k}},\label{eq:IC_L}\\
\mathcal{M}^{\mathcal{R}}_{\text{IC}}(i)=\frac{e}{N_{\textbf{k}}\hbar}\text{Im}\sum_{\textbf{k}}\sum_{vv'c}E_{v'\textbf{k}}\psi_{v\textbf{k}}(i)\psi^*_{v'\textbf{k}}(i)X^{\dagger}_{cv'\textbf{k}}Y_{vc\textbf{k}}.\label{eq:IC_R}
\end{gather} The combinations of these markers then yield the $\mathcal{M}^{\mathcal{C},\mathcal{C}}$, $\mathcal{M}^{\mathcal{E},\mathcal{E}}$, and $\mathcal{M}^{\text{lin}}$ markers.

\section{Microscopic tight-binding currents}
\label{AppB}

Here, we give further details on the
calculation of the microscopic current density in the TB representation. Our
derivation in Sec.~\ref{AvgCurr} was centered on continuum models, and we
briefly noted that in a TB context Eq.~(\ref{eq:ContCurr}) turns into a
discrete sum over TB sites.

Due to the discrete nature of TB models, microscopic currents are more
naturally defined along straight-line paths, which we refer to as bonds,
between different TB sites, rather than at individual TB sites. Assuming
that each TB site hosts a single orbital (the generalization to the
multi-orbital case being straightforward), the bond current between TB
sites $i$ and $j$ arises from the probabilities for a particle to hop
from site $i$ to site $j$ and vice-versa. Recall from the introduction to Sec.~\ref{Results} that we have adopted
a diagonal approximation to the matrix
elements of the position operators in the TB basis, \begin{equation}
\langle i|\hat{\textbf{r}}|j\rangle=\textbf{r}_i\delta_{ij}, \end{equation}
where $\textbf{r}_i$ denotes the position of TB orbital $i$. With this
approximation in mind, the matrix elements of the velocity operator
$\hat{\textbf{v}}=(1/i\hbar)[\hat{\textbf{r}},H]$ in the TB basis are expressed as
\begin{equation}
\langle i|\hat{\textbf{v}}|j\rangle
  =\frac{1}{i\hbar}(\textbf{r}_i-\textbf{r}_j)\langle i|H|j\rangle.
\label{eq:vme}
\end{equation}

Given a set of occupied states $|\psi_n\rangle$, the total current in the system is given by
\begin{equation}
\textbf{J}=\sum^{\text{occ}}_n\langle\psi_n|\hat{\textbf{J}}|\psi_n\rangle
\end{equation}
where $\hat{\textbf{J}}=e\hat{\textbf{v}}$ is the total current operator. Inserting resolutions of
the identity in terms of the TB sites on either side of $\hat{\textbf{J}}$, we
get
\begin{equation}
\textbf{J}=e\sum^{\text{occ}}_n\sum_{ij}\langle\psi_n|i\rangle\langle
i|\hat{\textbf{v}}|j\rangle\langle j|\psi_n\rangle
=e\sum_{ij}\langle j|P|i\rangle\langle i|\hat{\textbf{v}}|j\rangle
\end{equation}
where $P$ is the ground state projector.
Substituting Eq.~(\ref{eq:vme}), this can be written as
\begin{equation}
\textbf{J}=\sum_{\langle ij\rangle} \textbf{J}_{\langle ij\rangle}
\end{equation}
where the sum runs over distinct pairs
$\langle ij\rangle$ of sites and
\begin{equation}
\textbf{J}_{\langle ij\rangle} = \frac{2e}{\hbar} (\textbf{r}_i-\textbf{r}_j)
\text{Im}[\langle j|P|i\rangle\langle i|H|j\rangle]\label{eq:totalbond}
\end{equation}
describes the total current flowing on the bond $\langle ij\rangle$.

Since the current is assumed to flow uniformly along the straight-line
bond connecting the sites, it makes no difference if we treat
it as distributed uniformly along the length, concentrated at the
center, or partitioned between the two end points of the bond.
We make the last choice in order to write the total current as a sum
over sites, $\textbf{J}=\sum_i \textbf{j}_i$, where
\begin{equation}
\textbf{j}_i=\frac{1}{2}\sum_j \textbf{J}_{\langle ij\rangle}.
\end{equation}

We treat this as the discrete analog of Eq.~(\ref{eq:ContCurr}),
and we use it in the calculations of the macroscopic hinge currents
in our TB models by performing the coarse-graining procedure of
Sec.~\ref{AvgCurr} on the $\textbf{j}_i$.

Finally, we note that in the case of the extended rod geometries used to find the macroscopic hinge currents in the main text, the valence projector $P$ is written in
terms of the rod valence eigenstates $\psi_{vk}$ as $P=(1/N_k)\sum_{vk}|\psi_{vk}\rangle\langle\psi_{vk}|$, where the sum is over the $N_k$ points of the mesh spanning the 1D Brillouin zone. This expression for $P$ is then substituted into Eq.~(\ref{eq:totalbond}).

\section{Point-dipole-based magnetic quadrupole moment}
\label{AppC}

We now present a proof of Eq.~(\ref{eq:dip-nodip}) of the main text. Consider a collection of point magnetic dipoles $\textbf{m}_{\mu}$ located at positions $\textbf{r}_{\mu}$. We will enumerate the dipoles with Greek letters and use Latin letters to label Cartesian indices. The microscopic current distribution $\textbf{j}(\textbf{r})$ of the collection of dipoles is \begin{equation}
    \textbf{j}(\textbf{r})=\nabla\times\left(\sum_{\mu}\textbf{m}_{\mu}\delta(\textbf{r}-\textbf{r}_{\mu})\right)=-\sum_{\mu}\textbf{m}_{\mu}\times\nabla\delta(\textbf{r}-\textbf{r}_{\mu}).
\end{equation}
We then find that $\textbf{r}\times\textbf{j}(\textbf{r})$ in the definition of the current-based MQM (see Eq.~(\ref{eq:MQM})) is given by

\begin{equation}
(\textbf{r}\times\textbf{j})_i=\sum_\mu[ r_km_{\mu,k}\partial_i-m_{\mu,i}r_k\partial_k]
\delta(\textbf{r}-\textbf{r}_{\mu})
\end{equation}
in an implied sum notation.
Substituting this expression into Eq.~(\ref{eq:MQM}) and integrating
by parts, we find

\begin{align}
Q_{ij}&=\frac{1}{3}\int d\textbf{r}\,(\textbf{r}\times\textbf{j})_i\,r_j\nonumber\\
&=\frac{1}{3}\sum_{\mu}\int d\textbf{r}\left[m_{\mu,i}\partial_k(r_kr_j)-m_{\mu,k}\partial_i(r_kr_j)\right]\delta(\textbf{r}-\textbf{r}_{\mu})\nonumber\\
&=\sum_{\mu}m_{\mu,i}r_{\mu,j}-\frac{1}{3}\delta_{ij}m_{\mu,k}r_{\mu,k} \nonumber\\
&=Q^\text{dip}_{ij}-\frac{1}{3}\delta_{ij}Q_{kk}^\text{dip},
\end{align}
where we use the definition of $Q^{\text{dip}}_{ij}$ introduced in Eq.~(\ref{eq:Q_dip}). 
We have thus proved Eq.~(\ref{eq:dip-nodip}).

\section{Numerical TB model results}
\label{AppD}
Each of the TB models of Sec.~\ref{Results} features 2D layers coupled by interlayer couplings that are along the layer stacking direction. In going from one layer to the next, the couplings vary between two values denoted by $t_3$ and $t'_3$. When $t_3=t'_3$, the systems are found to exhibit axion-odd symmetries, and only axion-even symmetries otherwise. In each of the models, all parameters are kept fixed except for $t'_3$, which is used to bring the systems into and out of the axion-odd regime. For each model, $0.05\leq t'_3\leq0.15$. 

In the main text, we reported the values of the facet magnetizations $M_\perp^\text{B}$, $M_\perp^\text{F}$, and $M_\perp^\text{L}$ found from the $\mathcal{CC}$, $\mathcal{EE}$, and `lin' markers. Furthermore, we found the differences of facet magnetizations at different surfaces $\Delta M^\text{FB}$, $\Delta M^\text{BL}$, and $\Delta M^\text{LF}$, and compared them to the values of the directly computed macroscopic hinge currents. We did this for the parameter values $t'_3=0.1$ and $0.15$. In this Appendix, we report this data for $t'_3=0.05$, $0.075$, and $0.125$. The values of the magnetizations and currents are subject to the same symmetry constraints in the models' respective axion-even regimes that are described in the main text.

\subsection{Alternating Haldane layers}

\begin{table}[H]
\caption{Surface magnetizations and hinge currents
(in units of $10^{-5} e/\hbar$) for the alternating Haldane model
with $t'_3=0.05$ (axion-even).}
\begin{ruledtabular}
\begin{tabular}{lcccccc}
& $M_\perp^\text{B}$ & $M_\perp^\text{F}$ & $M_\perp^\text{L}$ &
$\Delta M^\text{FB}$ & $\Delta M^\text{BL}$ & $\Delta M^\text{LF}$ \\
\hline
$M^{\cal CC}$ & 1.8148 & $-$0.0493 & $-$0.0096 & $-$1.8641 & 1.8244 & 0.0397 \\ 
$M^{\cal EE}$ & 1.8148 & $-$0.0493 & $-$0.0096 & $-$1.8641 & 1.8244 & 0.0397 \\ 
$M^\text{lin}$ & 1.8148 & $-$0.0493 & $-$0.0096 & $-$1.8641 & 1.8244 & 0.0397 \\ 
$I_\text{calc}$ & & & & $-$1.8641 & 1.8244 & 0.0397
\end{tabular}
\end{ruledtabular}
\end{table}

\begin{table}[h]
\caption{Surface magnetizations and hinge currents
(in units of $10^{-5} e/\hbar$) for the alternating Haldane model
with $t'_3=0.075$ (axion-even).}
\begin{ruledtabular}
\begin{tabular}{lcccccc}
& $M_\perp^\text{B}$ & $M_\perp^\text{F}$ & $M_\perp^\text{L}$ &
$\Delta M^\text{FB}$ & $\Delta M^\text{BL}$ & $\Delta M^\text{LF}$ \\
\hline
$M^{\cal CC}$ & 1.8120 & $-$0.0288 & $-$0.0056 & $-$1.8408 & 1.8176 & 0.0232 \\ 
$M^{\cal EE}$ & 1.8120 & $-$0.0288 & $-$0.0056 & $-$1.8408 & 1.8176 & 0.0232 \\ 
$M^\text{lin}$ & 1.8120 & $-$0.0288 & $-$0.0056 & $-$1.8408 & 1.8176 & 0.0232 \\ 
$I_\text{calc}$ & & & & $-$1.8408 & 1.8176 & 0.0232
\end{tabular}
\end{ruledtabular}
\end{table}

\begin{table}[h]
\caption{Surface magnetizations and hinge currents
(in units of $10^{-5} e/\hbar$) for the alternating Haldane model
with $t'_3=0.125$ (axion-even).}
\begin{ruledtabular}
\begin{tabular}{lcccccc}
& $M_\perp^\text{B}$ & $M_\perp^\text{F}$ & $M_\perp^\text{L}$ &
$\Delta M^\text{FB}$ & $\Delta M^\text{BL}$ & $\Delta M^\text{LF}$ \\
\hline
$M^{\cal CC}$ & 1.8032 & 0.0372 & 0.0073 & $-$1.7659 & 1.7959 & $-$0.0299 \\ 
$M^{\cal EE}$ & 1.8032 & 0.0372 & 0.0073 & $-$1.7659 & 1.7959 & $-$0.0299 \\ 
$M^\text{lin}$ & 1.8032 & 0.0372 & 0.0073 & $-$1.7659 & 1.7959 & $-$0.0299 \\ 
$I_\text{calc}$ & & & & $-$1.7659 & 1.7959 & $-$0.0299
\end{tabular}
\end{ruledtabular}
\end{table}

\subsection{Two-layer plaquette model: straight edges}
\begin{table}[H]
\caption{Surface magnetizations and hinge currents (in units of $10^{-6} e/\hbar$) for the two-layer plaquette model with straight-edge terminations and with $t'_3=0.05$ (axion-even).}
\begin{ruledtabular}
\begin{tabular}{lcccccc}
& $M_\perp^\text{B}$ & $M_\perp^\text{F}$ & $M_\perp^\text{L}$ &
$\Delta M^\text{FB}$ & $\Delta M^\text{BL}$ & $\Delta M^\text{LF}$ \\
\hline
$M^{\cal CC}$ & 2.4215 & $-$0.0174 & $-$0.0174 & $-$2.4390 & 2.4390 & 0.0000 \\ 
$M^{\cal EE}$ & 2.4224 & $-$0.0179 & $-$0.0179 & $-$2.4403 & 2.4403 & 0.0000 \\ 
$M^\text{lin}$ & 2.4221 & $-$0.0178 & $-$0.0178 & $-$2.4399 & 2.4399 & 0.0000 \\ 
$I_\text{calc}$ & & & & $-$2.4399 & 2.4399 & 0.0000 
\end{tabular}
\end{ruledtabular}
\end{table}

\begin{table}[h]
\caption{Surface magnetizations and hinge currents (in units of $10^{-6} e/\hbar$) for the two-layer plaquette model with straight-edge terminations and with $t'_3=0.075$ (axion-even).}
\begin{ruledtabular}
\begin{tabular}{lcccccc}
& $M_\perp^\text{B}$ & $M_\perp^\text{F}$ & $M_\perp^\text{L}$ &
$\Delta M^\text{FB}$ & $\Delta M^\text{BL}$ & $\Delta M^\text{LF}$ \\
\hline
$M^{\cal CC}$ & 2.4030 & $-$0.0103 & $-$0.0103 & $-$2.4133 & 2.4133 & 0.0000 \\ 
$M^{\cal EE}$ & 2.4035 & $-$0.0106 & $-$0.0106 & $-$2.4141 & 2.4141 & 0.0000 \\ 
$M^\text{lin}$ & 2.4033 & $-$0.0105 & $-$0.0105 & $-$2.4138 & 2.4138 & 0.0000 \\ 
$I_\text{calc}$ & & & & $-$2.4138 & 2.4138 & 0.0000 
\end{tabular}
\end{ruledtabular}
\end{table}

\begin{table}[h]
\caption{Surface magnetizations and hinge currents (in units of $10^{-6} e/\hbar$) for the two-layer plaquette model with straight-edge terminations and with $t'_3=0.125$ (axion-even).}
\begin{ruledtabular}
\begin{tabular}{lcccccc}
& $M_\perp^\text{B}$ & $M_\perp^\text{F}$ & $M_\perp^\text{L}$ &
$\Delta M^\text{FB}$ & $\Delta M^\text{BL}$ & $\Delta M^\text{LF}$ \\
\hline
$M^{\cal CC}$ & 2.3430 & 0.0139 & 0.0139 & $-$2.3291 & 2.3291 & 0.0000 \\ 
$M^{\cal EE}$ & 2.3422 & 0.0143 & 0.0143 & $-$2.3280 & 2.3280 & 0.0000 \\ 
$M^\text{lin}$ & 2.3425 & 0.0141 & 0.0141 & $-$2.3284 & 2.3284 & 0.0000 \\ 
$I_\text{calc}$ & & & & $-$2.3284 & 2.3284 & 0.0000 
\end{tabular}
\end{ruledtabular}
\end{table}

\subsection{Two-layer plaquette model: zigzag edges}

\begin{table}[H]
\caption{Surface magnetizations and hinge currents (in units of $10^{-6} e/\hbar$) for the two-layer plaquette model with zigzag terminations and with $t'_3=0.05$ (axion-even).}
\begin{ruledtabular}
\begin{tabular}{lcccccc}
& $M_\perp^\text{B}$ & $M_\perp^\text{F}$ & $M_\perp^\text{L}$ &
$\Delta M^\text{FB}$ & $\Delta M^\text{BL}$ & $\Delta M^\text{LF}$ \\
\hline
$M^{\cal CC}$ & 2.4215 & 0.0011 & $-$0.0359 & $-$2.4204 & 2.4574 & $-$0.0370 \\ 
$M^{\cal EE}$ & 2.4224 & 0.3247 & $-$0.3605 & $-$2.0977 & 2.7829 & $-$0.6852 \\ 
$M^\text{lin}$ & 2.4221 & 0.2168 & $-$0.2523 & $-$2.2053 & 2.6744 & $-$0.4691 \\ 
$I_\text{calc}$ & & & & $-$2.2053 & 2.6744 & $-$0.4691 
\end{tabular}
\end{ruledtabular}
\end{table}

\begin{table}[h]
\caption{Surface magnetizations and hinge currents (in units of $10^{-6} e/\hbar$) for the two-layer plaquette model with zigzag terminations and with $t'_3=0.075$ (axion-even).}
\begin{ruledtabular}
\begin{tabular}{lcccccc}
& $M_\perp^\text{B}$ & $M_\perp^\text{F}$ & $M_\perp^\text{L}$ &
$\Delta M^\text{FB}$ & $\Delta M^\text{BL}$ & $\Delta M^\text{LF}$ \\
\hline
$M^{\cal CC}$ & 2.4030 & 0.0006 & $-$0.0212 & $-$2.4024 & 2.4242 & $-$0.0218 \\ 
$M^{\cal EE}$ & 2.4035 & 0.1909 & $-$0.2121 & $-$2.2126 & 2.6156 & $-$0.4030 \\ 
$M^\text{lin}$ & 2.4033 & 0.1275 & $-$0.1485 & $-$2.2758 & 2.5518 & $-$0.2760 \\ 
$I_\text{calc}$ & & & & $-$2.2758 & 2.5518 & $-$0.2760
\end{tabular}
\end{ruledtabular}
\end{table}

\begin{table}[h]
\caption{Surface magnetizations and hinge currents (in units of $10^{-6} e/\hbar$) for the two-layer plaquette model with zigzag terminations and with $t'_3=0.125$ (axion-even).}
\begin{ruledtabular}
\begin{tabular}{lcccccc}
& $M_\perp^\text{B}$ & $M_\perp^\text{F}$ & $M_\perp^\text{L}$ &
$\Delta M^\text{FB}$ & $\Delta M^\text{BL}$ & $\Delta M^\text{LF}$ \\
\hline
$M^{\cal CC}$ & 2.3430 & $-$0.0007 & 0.0284 & $-$2.3437 & 2.3145 & 0.0291 \\ 
$M^{\cal EE}$ & 2.3422 & $-$0.2518 & 0.2803 & $-$2.5941 & 2.0619 & 0.5322 \\ 
$M^\text{lin}$ & 2.3425 & $-$0.1681 & 0.1964 & $-$2.5106 & 2.1461 & 0.3645 \\ 
$I_\text{calc}$ & & & & $-$2.5106 & 2.1461 & 0.3645 
\end{tabular}
\end{ruledtabular}
\end{table}

\subsection{Four-layer plaquette model: straight edges}

\begin{table}[H]
\caption{Surface magnetizations and hinge currents (in units of $10^{-7} e/\hbar$) for the four-layer plaquette model with straight-edge terminations and with $t'_3=0.05$ (axion-even).}
\begin{ruledtabular}
\begin{tabular}{lcccccc}
& $M_\perp^\text{B}$ & $M_\perp^\text{F}$ & $M_\perp^\text{L}$ &
$\Delta M^\text{FB}$ & $\Delta M^\text{BL}$ & $\Delta M^\text{LF}$ \\
\hline
$M^{\cal CC}$ & 3.26720 & $-$0.00555 & $-$0.00627 & $-$3.27276 & 3.27348 & $-$0.00072 \\ 
$M^{\cal EE}$ & 3.26732 & $-$0.00557 & $-$0.00638 & $-$3.27289 & 3.27371 & $-$0.00082 \\ 
$M^\text{lin}$ & 3.26728 & $-$0.00556 & $-$0.00635 & $-$3.27284 & 3.27363 & $-$0.00078 \\ 
$I_\text{calc}$ & & & & $-$3.27285 & 3.27363 & $-$0.00078
\end{tabular}
\end{ruledtabular}
\end{table}

\begin{table}[h]
\caption{Surface magnetizations and hinge currents (in units of $10^{-7} e/\hbar$) for the four-layer plaquette model with straight-edge terminations and with $t'_3=0.075$ (axion-even).}
\begin{ruledtabular}
\begin{tabular}{lcccccc}
& $M_\perp^\text{B}$ & $M_\perp^\text{F}$ & $M_\perp^\text{L}$ &
$\Delta M^\text{FB}$ & $\Delta M^\text{BL}$ & $\Delta M^\text{LF}$ \\
\hline
$M^{\cal CC}$ & 3.25997 & $-$0.00301 & $-$0.00392 & $-$3.26298 & 3.26388 & $-$0.00091 \\ 
$M^{\cal EE}$ & 3.26004 & $-$0.00298 & $-$0.00401 & $-$3.26302 & 3.26405 & $-$0.00103 \\ 
$M^\text{lin}$ & 3.26002 & $-$0.00299 & $-$0.00398 & $-$3.26301 & 3.26400 & $-$0.00099 \\ 
$I_\text{calc}$ & & & & $-$3.26301 & 3.26400 & $-$0.00099
\end{tabular}
\end{ruledtabular}
\end{table}

\begin{table}[h]
\caption{Surface magnetizations and hinge currents (in units of $10^{-7} e/\hbar$) for the four-layer plaquette model with straight-edge terminations and with $t'_3=0.125$ (axion-even).}
\begin{ruledtabular}
\begin{tabular}{lcccccc}
& $M_\perp^\text{B}$ & $M_\perp^\text{F}$ & $M_\perp^\text{L}$ &
$\Delta M^\text{FB}$ & $\Delta M^\text{BL}$ & $\Delta M^\text{LF}$ \\
\hline
$M^{\cal CC}$ & 3.23673 & 0.00527 & 0.00375 & $-$3.23146 & 3.23298 & $-$0.00152 \\ 
$M^{\cal EE}$ & 3.23664 & 0.00541 & 0.00369 & $-$3.23122 & 3.23294 & $-$0.00172 \\ 
$M^\text{lin}$ & 3.23667 & 0.00536 & 0.00371 & $-$3.23130 & 3.23296 & $-$0.00165 \\ 
$I_\text{calc}$ & & & & $-$3.23130 & 3.23296 & $-$0.00165
\end{tabular}
\end{ruledtabular}
\end{table}

\subsection{Four-layer plaquette model: zigzag edges}

\begin{table}[H]
\caption{Surface magnetizations and hinge currents (in units of $10^{-7} e/\hbar$) for the four-layer plaquette model with zigzag terminations and with $t'_3=0.05$ (axion-even).}
\begin{ruledtabular}
\begin{tabular}{lcccccc}
& $M_\perp^\text{B}$ & $M_\perp^\text{F}$ & $M_\perp^\text{L}$ &
$\Delta M^\text{FB}$ & $\Delta M^\text{BL}$ & $\Delta M^\text{LF}$ \\
\hline
$M^{\cal CC}$ & 3.26720 & $-$0.00599 & $-$0.00599 & $-$3.27319 & 3.27319 & 0.00000 \\ 
$M^{\cal EE}$ & 3.26732 & $-$0.00605 & $-$0.00605 & $-$3.27337 & 3.27337 & 0.00000 \\ 
$M^\text{lin}$ & 3.26728 & $-$0.00603 & $-$0.00603 & $-$3.27331 & 3.27331 & 0.00000 \\ 
$I_\text{calc}$ & & & & $-$3.27331 & 3.27331 & 0.00000
\end{tabular}
\end{ruledtabular}
\end{table}

\begin{table}[h]
\caption{Surface magnetizations and hinge currents (in units of $10^{-7} e/\hbar$) for the four-layer plaquette model with zigzag terminations and with $t'_3=0.075$ (axion-even).}
\begin{ruledtabular}
\begin{tabular}{lcccccc}
& $M_\perp^\text{B}$ & $M_\perp^\text{F}$ & $M_\perp^\text{L}$ &
$\Delta M^\text{FB}$ & $\Delta M^\text{BL}$ & $\Delta M^\text{LF}$ \\
\hline
$M^{\cal CC}$ & 3.25997 & $-$0.00351 & $-$0.00351 & $-$3.26347 & 3.26347 & 0.00000 \\ 
$M^{\cal EE}$ & 3.26004 & $-$0.00354 & $-$0.00354 & $-$3.26358 & 3.26358 & 0.00000 \\ 
$M^\text{lin}$ & 3.26002 & $-$0.00353 & $-$0.00353 & $-$3.26355 & 3.26355 & 0.00000 \\ 
$I_\text{calc}$ & & & & $-$3.26355 & 3.26355 & 0.00000
\end{tabular}
\end{ruledtabular}
\end{table}

\begin{table}[h]
\caption{Surface magnetizations and hinge currents (in units of $10^{-7} e/\hbar$) for the four-layer plaquette model with zigzag terminations and with $t'_3=0.125$ (axion-even).}
\begin{ruledtabular}
\begin{tabular}{lcccccc}
& $M_\perp^\text{B}$ & $M_\perp^\text{F}$ & $M_\perp^\text{L}$ &
$\Delta M^\text{FB}$ & $\Delta M^\text{BL}$ & $\Delta M^\text{LF}$ \\
\hline
$M^{\cal CC}$ & 3.23673 & 0.00456 & 0.00456 & $-$3.23217 & 3.23217 & 0.00000 \\ 
$M^{\cal EE}$ & 3.23664 & 0.00461 & 0.00461 & $-$3.23203 & 3.23203 & 0.00000 \\ 
$M^\text{lin}$ & 3.23667 & 0.00459 & 0.00459 & $-$3.23207 & 3.23207 & 0.00000 \\ 
$I_\text{calc}$ & & & & $-$3.23207 & 3.23207 & 0.00000
\end{tabular}
\end{ruledtabular}
\end{table}

\bibliography{pap}

\end{document}